\documentclass[twocolumn]{aastex631}

\usepackage{graphicx}
\usepackage{amsmath}
\usepackage{inputenc}
\usepackage[encapsulated]{CJK}
\usepackage{ucs}
\usepackage{chngcntr}
\graphicspath{{./}{figs/}}

\newcommand{\ruby}{RUBIES-EGS-}

\newcommand{\prospector}{\texttt{Prospector}}

\newcommand{\msun}{{\rm M}_{\odot}}
\newcommand{\zsun}{{\rm Z}_{\odot}}

\newcommand{\mstar}{M_\star}
\newcommand{\mstarmbh}{M_\star-M_{\rm BH}}

\newcommand{\zspec}{z_{\rm spec}}

\newcommand{\ha}{H$\alpha$}
\newcommand{\hb}{H$\beta$}

\newcommand{\nii}{[N\,{\sc ii}]}
\newcommand{\oii}{[O\,{\sc ii}]}
\newcommand{\oiii}{[O\,{\sc iii}]}

\newcommand{\neiii}{[Ne\,{\sc iii}]}
\newcommand{\kms}{\rm km\,s^{-1}}

\shorttitle{Balmer Breaks at $\zspec = 6.7-8.4$}
\shortauthors{Wang et al}

\begin{document}

\title{RUBIES: Evolved Stellar Populations with Extended Formation Histories at $z \sim 7-8$ in Candidate Massive Galaxies Identified with JWST/NIRSpec}

\correspondingauthor{Bingjie Wang}
\email{bwang@psu.edu}

\author[0000-0001-9269-5046]{Bingjie Wang (\begin{CJK*}{UTF8}{gbsn}王冰洁\ignorespacesafterend\end{CJK*})}
\affiliation{Department of Astronomy \& Astrophysics, The Pennsylvania State University, University Park, PA 16802, USA}
\affiliation{Institute for Computational \& Data Sciences, The Pennsylvania State University, University Park, PA 16802, USA}
\affiliation{Institute for Gravitation and the Cosmos, The Pennsylvania State University, University Park, PA 16802, USA}

\author[0000-0001-6755-1315]{Joel Leja}
\affiliation{Department of Astronomy \& Astrophysics, The Pennsylvania State University, University Park, PA 16802, USA}
\affiliation{Institute for Computational \& Data Sciences, The Pennsylvania State University, University Park, PA 16802, USA}
\affiliation{Institute for Gravitation and the Cosmos, The Pennsylvania State University, University Park, PA 16802, USA}

\author[0000-0002-2380-9801]{Anna de Graaff}
\affiliation{Max-Planck-Institut f\"ur Astronomie, K\"onigstuhl 17, D-69117, Heidelberg, Germany}

\author[0000-0003-2680-005X]{Gabriel B. Brammer}
\affiliation{Cosmic Dawn Center (DAWN), Copenhagen, Denmark}
\affiliation{Niels Bohr Institute, University of Copenhagen, Jagtvej 128, Copenhagen, Denmark}

\author[0000-0001-8928-4465]{Andrea Weibel}
\affiliation{Department of Astronomy, University of Geneva, Chemin Pegasi 51, 1290 Versoix, Switzerland}

\author[0000-0002-8282-9888]{Pieter van Dokkum}
\affiliation{Department of Astronomy, Yale University, New Haven, CT 06511, USA}

\author[0009-0005-2295-7246]{Josephine F.W. Baggen}
\affiliation{Department of Astronomy, Yale University, New Haven, CT 06511, USA}

\author[0000-0002-1714-1905]{Katherine A. Suess}
\thanks{NHFP Hubble Fellow}
\affiliation{Kavli Institute for Particle Astrophysics and Cosmology and Department of Physics, Stanford University, Stanford, CA 94305, USA}

\author[0000-0002-5612-3427]{Jenny E. Greene}
\affiliation{Department of Astrophysical Sciences, Princeton University, Princeton, NJ 08544, USA}

\author[0000-0001-5063-8254]{Rachel Bezanson}
\affiliation{Department of Physics \& Astronomy and PITT PACC, University of Pittsburgh, Pittsburgh, PA 15260, USA}

\author[0000-0001-7151-009X]{Nikko J. Cleri}
\affiliation{Department of Physics and Astronomy, Texas A\&M University, College Station, TX, 77843-4242 USA}
\affiliation{George P.\ and Cynthia Woods Mitchell Institute for Fundamental Physics and Astronomy, Texas A\&M University, College Station, TX, 77843-4242 USA}

\author[0000-0002-3301-3321]{Michaela Hirschmann}
\affiliation{Institute of Physics, Lab for Galaxy Evolution, EPFL, Observatoire de Sauverny, Chemin Pegasi 51, 1290 Versoix, Switzerland}

\author[0000-0002-2057-5376]{Ivo Labb\'e}
\affiliation{Centre for Astrophysics and Supercomputing, Swinburne University of Technology, Melbourne, VIC 3122, Australia}

\author[0000-0003-2871-127X]{Jorryt Matthee}
\affiliation{Institute of Science and Technology Austria (ISTA), Am Campus 1, 3400 Klosterneuburg, Austria}

\author[0000-0002-2446-8770]{Ian McConachie}
\affiliation{Department of Physics and Astronomy, University of California, Riverside, Riverside, CA 92521, USA}

\author[0000-0003-2895-6218]{Rohan P. Naidu}
\thanks{NHFP Hubble Fellow}
\affiliation{MIT Kavli Institute for Astrophysics and Space Research, Cambridge, MA 02139, USA}

\author[0000-0002-7524-374X]{Erica Nelson}
\affiliation{Department for Astrophysical and Planetary Science, University of Colorado, Boulder, CO 80309, USA}

\author[0000-0001-5851-6649]{Pascal A. Oesch}
\affiliation{Department of Astronomy, University of Geneva, Chemin Pegasi 51, 1290 Versoix, Switzerland}
\affiliation{Cosmic Dawn Center (DAWN), Copenhagen, Denmark}

\author[0000-0003-4075-7393]{David J. Setton}\thanks{Brinson Prize Fellow}
\affiliation{Department of Astrophysical Sciences, Princeton University, Princeton, NJ 08544, USA}

\author[0000-0003-2919-7495]{Christina C. Williams}
\affiliation{NSF's National Optical-Infrared Astronomy Research Laboratory, Tucson, AZ 85719, USA}
\affiliation{Steward Observatory, University of Arizona, Tucson, AZ 85721, USA}

\begin{abstract}

The identification of red, apparently massive galaxies at $z>7$ in early James Webb Space Telescope (JWST) photometry suggests a strongly accelerated timeline compared to standard models of galaxy growth. A major uncertainty in the interpretation is whether the red colors are caused by evolved stellar populations, dust, or other effects such as emission lines or AGN. Here we show that three of the massive galaxy candidates at $z=6.7-8.4$ have prominent Balmer breaks in JWST/NIRSpec spectroscopy from the RUBIES program. The Balmer breaks demonstrate unambiguously that stellar emission dominates at $\lambda_{\rm rest} = 0.4\,\mu$m, and require formation histories extending hundreds of Myr into the past in galaxies only 600--800 Myr after the Big Bang. Two of the three galaxies also show broad Balmer lines, with H$\beta$ FWHM $>2500~{\rm km\,s^{-1}}$, suggesting that dust-reddened AGN contribute to, or even dominate, the SEDs of these galaxies at $\lambda_{\rm rest}\gtrsim 0.6\,\mu$m. 
All three galaxies have relatively narrow [O\,\textsc{iii}] lines, seemingly ruling out a high-mass interpretation if the lines arise in dynamically-relaxed, inclined disks.
Yet, the inferred masses also remain highly uncertain. We model the high-quality spectra using Prospector to decompose the continuum into stellar and AGN components, and explore limiting cases in stellar/AGN contribution. This produces a wide range of possible stellar masses, spanning $M_\star \sim 10^9 - 10^{11}~{\rm M_{\odot}}$. Nevertheless, all fits suggest a very early and rapid formation, most of which follow with a truncation in star formation. Potential origins and evolutionary tracks for these objects are  discussed, from the cores of massive galaxies to low-mass galaxies with over-massive black holes. Intriguingly, we find all of these explanations to be incomplete; deeper and redder data are needed to understand the physics of these systems.

\end{abstract}


\section{Introduction}

In the cores of the most massive galaxies in the local universe, stars have inferred stellar age of $\sim13$ Gyr and high $\alpha$-element abundance, suggesting their stellar components are formed at $z\gtrsim5$ in a spectacular and short burst of star formation (e.g., \citealt{thomas05}). This hypothesis was bolstered by the discovery of their putative $z\sim2$ progenitors, compact galaxies with high stellar masses $\sim10^{11}$ M$_{\odot}$ and small effective radii of $\sim1$~kpc, which importantly have stellar densities similar to the cores of $z\sim0$ ellipticals \citep{bezanson2009}. When exactly their stellar bodies formed, however, has not yet been clearly established. While some likely compact star-forming progenitors have been identified at $z=2-3$ \citep{nelson14,barro14}, simulations and number density arguments suggest that at least some of these massive cores must have formed earlier \citep{wellons15}. This has since been buttressed by the James Webb Space Telescope (JWST) discovering and characterizing massive quiescent galaxies at $z=2-5$, with stellar bodies of $10^{10.5-11}$ M$_{\odot}$ and inferred formation redshifts of $z\gtrsim7$ \citep{Carnall2023,Glazebrook2023,deGraaff2024,Park2024}. Taken at face value, this implies a very rapid formation and quenching of very massive galaxies in the first billion years of the universe---a spectacular event that should produce a huge amount of observable light. Yet, while candidates have been found (e.g., \citealt{Hashimoto2018,Williams2023}), the progenitors of these quenched galaxies existing at $t_{\mathrm{univ}} \sim 1$ Gyr have yet to be conclusively identified.

Contemporaneously, one of the most surprising early discoveries made with the JWST is the identification of a population of seemingly massive ($\mstar \gtrsim 10^{10}~\msun$), compact (effective radii $\lesssim 1$ kpc), and rest-optical red galaxies at redshift $z>6$ via a double-break color selection (\citealt{Labbe2023:massive}; hereafter L23). This selection targets distinct spectral energy distributions (SEDs) which include a drop-out at $\sim 1~\mu$m and a very red color at $\sim 3~\mu$m in observed frame. This is a highly efficient selection for $7 \lesssim z \lesssim 9$ objects, but the red rest-frame optical color may be driven by very different underlying physics: it could come from a Balmer break, from strong emission lines, from a very red continuum, or from some combination of these features.

In their main analysis, L23 interpreted the red color as a combination of a Balmer break and a very red rest-optical stellar continuum. This interpretation yields very high stellar mass-to-light ratios (M/L) and implies extreme stellar masses, up to $\mstar \sim 10^{11}~\msun$. While massive galaxies must form early and quickly, this would imply that these objects both emerged earlier and hosted more mass than expected. Indeed, it soon became clear that such early massive galaxies are difficult to make in the standard model of cosmology, as the amount of baryons inferred in stars is comparable to the cosmic baryon abundance in these early halos \citep{Boylan-Kolchin2023}. Star formation feedback processes typically limit the fraction of baryons locked up in stars to far below the cosmic baryon fraction.

These uncomfortably high stellar masses prompted a wave of alternative explanations for the observed fluxes and colors, including extreme emission line galaxies \citep{Endsley2023}, a top-heavy stellar initial mass function \citep{Boylan-Kolchin2023,Steinhardt2023}, or obscured active galactic nuclei (AGN) \citep{Kocevski2023,Barro2024}. Obscured AGNs significantly lower the inferred stellar masses by contributing to the red optical continuum flux. They can also lower the photometric redshift, and so decrease the inferred number densities of massive galaxies at high redshifts. In addition to their red color, the L23 sample exhibits compact morphology, supporting the idea that AGN could contribute to the fluxes. 

Indeed, one of the bright red objects in L23 has been confirmed to host a broad-line AGN at a lower redshift of $z = 5.62$ \citep{Kocevski2023}, and the more broadly defined red compact sources (often dubbed as little red dots, or LRDs; e.g., \citealt{Labbe2023:agn,Matthee2023}) have shown a prevalence of broad-line AGNs \citep{Greene2023}. However, as found in \citet{Baggen2023}, most of the L23 sample are marginally resolved at rest-UV wavelengths---so their rest-UV flux at least is not dominated by a point source. Furthermore, the inferred stellar densities, assuming that the L23 objects are galaxies, are consistent with the central regions of today's elliptical galaxies.

Therefore, these objects may still be massive galaxies with evolved populations at high redshifts---in fact, it would be surprising if star formation slowed sufficiently to show prominent Balmer breaks at high redshift without influence from an AGN. The two massive quiescent systems currently known at $z\sim5$ both show indications of AGN activity \citep{Carnall2023,deGraaff2024}. It has long been suspected that the stellar cores of galaxies must have co-formed with their supermassive black holes; such a formation scenario is the simplest interpretation of the observed tight correlation between the mass of the bulge and the mass of the central supermassive black hole \citep{ferrarese00}. The key questions regarding these red objects are therefore three-fold: (i) are the photometric redshifts accurate; or more generally, what is the source of the red colors---emission lines, continua, or spectral breaks? (ii) do these objects host evolved stellar populations? and (iii) how much of the continuum is powered by stellar versus AGN emission? These core questions are unanswerable without rest-frame optical spectra.

Here we conduct a follow-up study on double-break candidates selected from the RUBIES JWST/NIRSpec spectroscopic program (JWST-GO-4233; PIs de Graaff \& Brammer). These targets partially overlap with the L23 sample, and were observed with high priority for their red colors (F150W--F444W $>$ 2), and bright apparent magnitudes (F444W $<$ 26 mag).
In this paper, we present three objects with detected Balmer breaks, existing as early as $\zspec=8.35$, suggesting a formation history extending hundreds of Myr into the past in galaxies only 600--800 Myr after the Big Bang. Unambiguous broad emission lines are also observed in 2/3 objects, motivating a deeper look at the source of the red continua.

The structure of this paper is as follows. Section~\ref{sec:data} provides an overview of the data. 
Section~\ref{sec:balmer_break} presents the key observational features of these objects, including observed Balmer breaks and broad lines.
Section~\ref{sec:lines} focuses on the analysis of the emission lines, while Section~\ref{sec:prosp} details the AGN and host galaxy composite SED modeling. 
Section~\ref{sec:res} presents evidence for/against the different proposed physical models, including kinematics, formation histories, and population-level characteristics.
We conclude in Section~\ref{sec:concl} with discussion of possible interpretations and evolutionary scenarios, and outstanding questions.

Where applicable, we adopt the best-fit cosmological parameters from the WMAP 9 yr results: $H_{0}=69.32$ ${\rm km \,s^{-1} \,Mpc^{-1}}$, $\Omega_{M}=0.2865$, and $\Omega_{\Lambda}=0.7135$ \citep{Hinshaw2013}. Unless otherwise mentioned, we assume the \citet{Chabrier2003} initial mass function, and report the median of the posterior, with associated 1$\sigma$ error bars being the 16th and 84th percentiles.

\section{Data\label{sec:data}}

\begin{figure*} 
\gridline{
  \fig{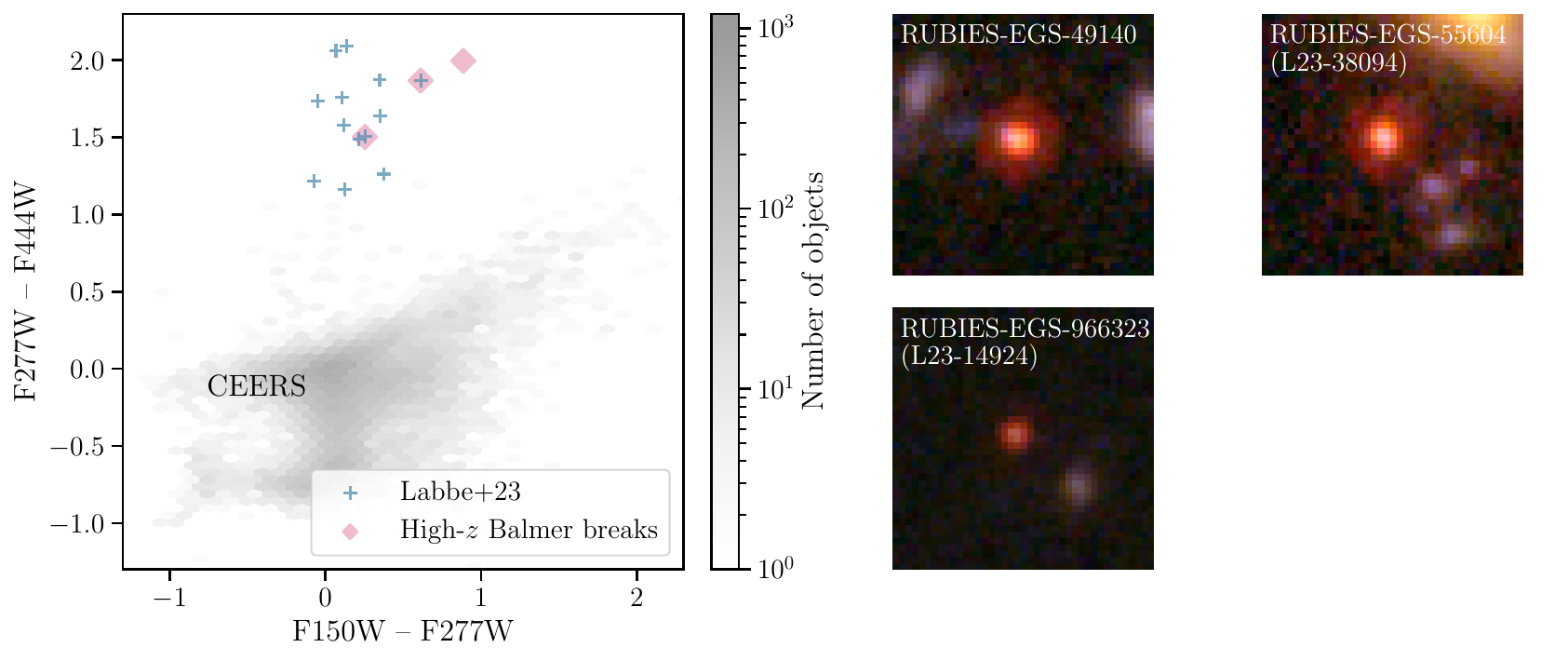}{0.95\textwidth}{}
}
\caption{Sample of this paper. 
(Left) This work focuses on continuum-detected sources with Balmer breaks, shown as solid diamonds on the color plane.
The L23 sample is denoted as plus signs, and all sources in CEERS brighter than 27 AB mag are grey hexagons included for reference.
(Right) We show the $2\times2$\arcsec\ color images of the three Balmer-break-detected sources, colors from JWST/NIRCam F115W, F277W, and F444W. They are remarkably compact at red wavelengths, with some evidence for spatial structure at blue wavelengths.
}
\label{fig:samp}
\end{figure*}

The spectroscopic survey RUBIES uses JWST/NIRSpec to observe approximately 4000--5000 sources selected from public NIRCam imaging in the EGS and UDS fields. It utilizes the NIRSpec micro-shutter array \citep[MSA;][]{Ferruit2022} with both the low-resolution Prism/CLEAR and medium-resolution G395M/F290LP disperser/filter combinations.
The sample in this paper was targeted in 6 masks observed in March 2024. Figure~\ref{fig:samp} presents an overview of the sample in color space.

Reduction of the imaging data is presented in \citet{Valentino2023} while the spectroscopic reductions are described in \citet{Heintz2024,Wang2024:brd}. Full details of the RUBIES observing program and data reduction will be detailed in A. de Graaff et al. in prep. The subsequent sections provide a brief summary.

\subsection{Imaging\label{subsec:img}}

The RUBIES targets in the EGS were selected from public JWST/NIRCam data from the Cosmic Evolution Early Release Science Survey (CEERS, JWST-GO-1345; PI Finkelstein; \citealt{Bagley2023,Finkelstein2023}), which provides imaging in 7 bands (F115W, F150W, F200W, F277W, F356W, F410M, and F444W).
Additionally, we use archival imaging in 7 different filters (F435W, F606W, F814W, F105W, F125W, F140W, and F160W) from the Hubble Space Telescope (HST) from the CANDELS survey \citep{Grogin2011,Koekemoer2011}. 

The image mosaics, with a pixel scale of $0.04\arcsec\,{\rm pix}^{-1}$, are publicly available in the DAWN JWST Archive (DJA; version 7.2) and were reduced using \texttt{grizli} \citep{grizli}. Fluxes are measured from PSF-matched images in circular apertures with a radius of $0.16\arcsec$, and then Kron-corrected to the total flux, as described in \citet{Weibel2024}.

\subsection{Spectroscopy and Sample Selection\label{subsec:spec}}

Each target was observed for 48~min in the Prism/CLEAR mode and the G395M/F290LP mode, using a standard 3-shutter slitlet and 3-point nodding pattern. The spectra are reduced, combined and extracted using the JWST calibration pipeline version 1.12.5 \citep{Backhaus2024} and \texttt{msaexp} \citep{Brammer2022}, corresponding to the version 2 reduction on DJA. To account for wavelength-dependent flux calibration that is not yet captured well by the pipeline, we re-normalize the Prism spectrum to match the NIRCam photometry using a dynamic high-order polynomial as described in Section~\ref{sec:prosp}. The G395M spectrum is subsequently rescaled by this polynomial. We find that there is a small systematic offset ($\approx 10-20\%$) in the flux calibration between the Prism and G395M spectra, which has recently also been reported by \citet{JADES2024}. To determine the offset we fit the \oiii\ doublet in both the Prism and G395M spectra (\S\,\ref{sec:lines}), and hence rescale the full G395M spectrum by the ratio of the two to match the flux calibration of the Prism spectrum.

This paper focuses on the 3 targets in this sample that exhibit clear Balmer breaks at $\zspec = 6.6 - 8.4$, found via visual inspection of the 2D and 1D spectra. These objects are extremely red in F277W -- F444W (Figure~\ref{fig:samp}), the characterization of red objects being one of the core targeting criteria of RUBIES.
In what follows, we establish the detections of Balmer breaks, and the broad emission lines, which are the two key characteristics of this sample.

\section{The Coexistence of Balmer Breaks and Broad Emission Lines at $\lowercase{z} = 6.6 - 8.4$\label{sec:balmer_break}}

\begin{figure*} 
\gridline{
  \fig{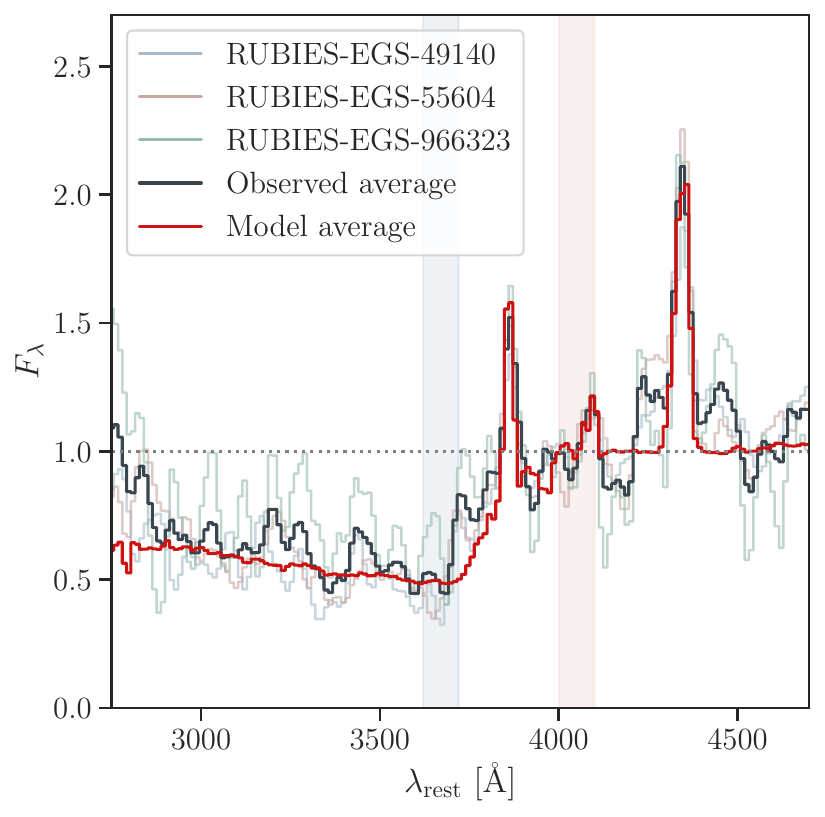}{0.45\textwidth}{(a)}
  \fig{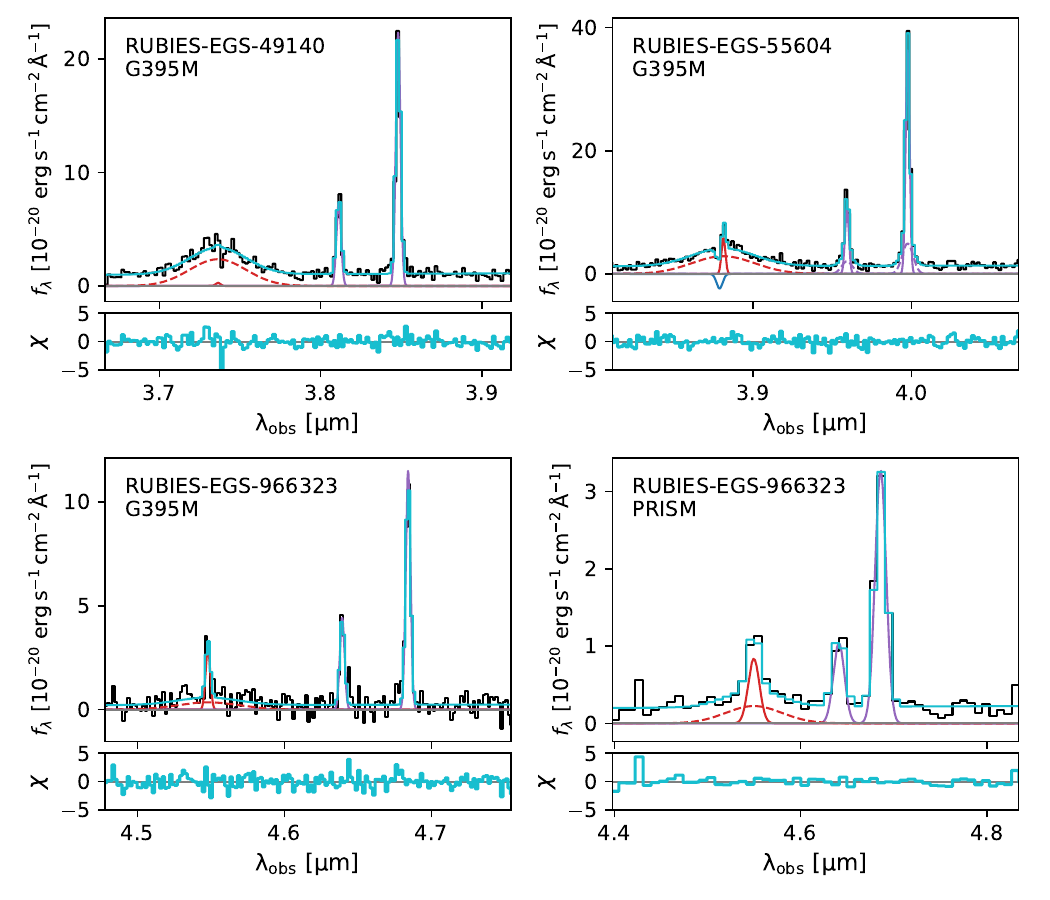}{0.52\textwidth}{(b)}
}
\caption{Characteristics of the sample of this paper. 
(a) Balmer breaks detected at $\zspec = 6.7 - 8.4$. The scaled individual spectra which exhibit potential Balmer breaks are plotted in light colors, and the averaged spectrum is shown in black. Over-plotted in red is the averaged best-fit model spectrum (\S\,\ref{sec:prosp}). All breaks are located at the expected wavelength, and show similar shapes, supporting the Balmer break interpretation. The break strength for the average observed spectrum, quantified by the flux ratio in the wavelength windows illustrated in blue and red shades, is 2.0.
(b) Emission line fits to the \hb\ and \oiii\ complex (\S\,\ref{sec:lines}). Unambiguous broadened \hb\ lines are found in the G395M spectrum for two sources (\ruby49140 and 55604), indicative of an AGN. A broad component of \hb\ is only marginally detected in both the G395M and Prism fits for \ruby966323.}
\label{fig:data}
\end{figure*}

The most striking discovery is the seeming appearance of Balmer breaks observed between $z=6.7-8.4$. These are produced in the atmospheres of older stars and typically only appear in evolved stellar populations after a significant reduction in star formation rate (SFR) lasting at least $\sim100$ Myr \citep{Bruzual1983,Hamilton1985,Worthey1994,Balogh1999}. It is surprising to see SEDs dominated by evolved stellar populations at these times---the age of the universe for the highest redshift object ($z=8.35$) is 610~Myr, giving very little time to form the stellar populations. This suggests an extremely rapid and early formation for these objects. The highest redshift candidate at $z=8.35$ represents the highest-redshift Balmer break identified to date, with the next-highest redshift being a $z=7.3$ object with a low mass of $10^{8.5}$ M$_{\odot}$ \citep{Looser2023}.

To verify that the observed spectral breaks are indeed Balmer breaks, as opposed to observational artifacts or a mis-identification of spectral breaks driven by other physics, we take the average of the three observed spectra after scaling each spectrum by the median of the continuum flux close to the break (between 4150 and 4250\,\AA\ in rest frame). The same average is then calculated for the best-fit galaxy model spectra (\S\,\ref{sec:prosp}). We show the normalized individual Prism spectra, over-plotted with the averaged model spectra in Figure~\ref{fig:data}. It is evident that all the breaks are located at the expected wavelength, and have similar shapes, buttressing the Balmer break interpretation. Critically, the stacked galaxy model spectra agree well with the data, suggesting that the model fits are also properly interpreting the light as evolved stellar populations. Section~\ref{sec:prosp} elaborates on the SED modeling.

A commonly used measure of the Balmer break strength is $D_{4000}$, as originally defined in \citet{Bruzual1983,Balogh1999}. The wavelength windows in these definitions locate redward of the Balmer limit (3645\,\AA), as they are intended for measuring the 4000\,\AA\ break. However, we are more interested in the Balmer break. We thus define a spectral break strength, taking the median of the fluxes in two windows at [3620,3720]\,\AA\ and [4000,4100]\,\AA\ instead. These windows avoid contamination from strong nebular line emission and are close enough to be minimally impacted by dust attenuation, or the overall slope of the spectrum. Our spectral break strength differs primarily from $D_{4000}$ for galaxies with ages $\lesssim 1.5$~Gyr, where the Balmer break is prominent but the 4000\,\AA\ break is less visible. It is therefore a better age indicator for high-redshift galaxies. A comparison between our new index with the \citet{Balogh1999} definition is supplemented in Appendix~\ref{app:d4000}. Meanwhile, the break strengths for \ruby49140, 55604, and 966323 are $2.44\pm0.10$, $2.18\pm0.11$, and $1.96\pm0.14$, respectively.

In addition to hosting Balmer breaks, we note that the broad emission lines are evident in the spectra of \ruby49140 and 55604, as shown in Figure~\ref{fig:data}, while the existence of broad \hb\ is more ambiguous for the final object \ruby966323. These observations motivate detailed emission line fitting to decompose them into broad and (potentially) narrow components, and also present an opportunity to estimate black hole masses. We do so in Section~\ref{sec:lines}.

\section{Emission Lines\label{sec:lines}}

\subsection{Emission Line Decomposition\label{subsec:lines}}

The emission line widths of the Balmer \hb\ and \oiii\ lines are modeled using the G395M spectra.
Prior to fitting we rescale the error spectrum, as we find that the noise estimated by the \texttt{msaexp} pipeline is underestimated compared to the observed pixel-to-pixel variation. 
To estimate the rescaling factor we select the region outside of the \hb\ and \oiii\ emission lines and calculate the ratio between the pixel-to-pixel variance and the median of the error spectrum; this results in a rescaling factor in the range $1.3-2.0$.

Our fitting methodology broadly follows that described in \citet{Wang2024:brd}: given the compact morphology of the sources \citep{Baggen2023}, we use a point source line-spread function (LSF) from \citet{deGraaff2023}, and the Markov Chain Monte Carlo (MCMC) ensemble sampler \texttt{emcee} \citep{emcee} to estimate the posteriors. Crucially, our method explicitly includes a systematic uncertainty on the model LSF, and therefore provides realistic measurement uncertainties for marginally resolved lines. 

We select the region around the emission line complex ($\pm0.2\,\micron$), and begin by masking the \hb\ line (a region of $7000\,\kms$) to fit only the \oiii\ doublet to estimate the narrow line width and redshift of the source. We use a single Gaussian component, the dispersion of which is the same for both lines of the \oiii\ doublet and constrained to be in the range $\rm \sigma_{\rm narrow}\in[0,500]~\kms$ using a uniform prior. The continuum is fit with a 1st order polynomial. We then explore whether the \oiii\ lines show evidence for a second broader component by fitting a 2-component Gaussian model, where the width of the second Gaussian is constrained to be larger than that of narrow component. The Bayesian information criterion (BIC) is computed to compare the two models \citep{jeffreys1961theory,Liddle2004}. We find evidence ($\rm \Delta BIC = 13$) for a broader component in the \oiii\ line for \ruby55604 with a dispersion $\sigma_{\rm broad, OIII} = 248_{-41}^{+61}\,\kms$, and no evidence for a second component in the other two sources.

Next, we include the \hb\ line in the fitting. We first fit a single Gaussian component for the \hb\ line. Given the limited S/N of the data, the width of this component is tied to that of the \oiii\ lines. We then include a second broad component for the \hb\ line, where $\sigma_{\rm broad}\in[500,2500]~\kms$. For \ruby55604 we find a blue-shifted absorption feature in the \hb\ line, which is modeled as an additional Gaussian component with a velocity offset $\Delta v \in[0,1000]~\kms$ and dispersion $\rm \sigma_{\rm outflow}\in[0,1000]~\kms$. Such absorption feature in Balmer and/or He lines has been seen in recent studies of AGNs as well \citep{Kocevski2024,Matthee2023,Wang2024:brd}. The multi-component fits and the residuals are shown in Figure~\ref{fig:data}.

Again, using the BIC to compare the models, we find that \ruby49140 and 55604 have unambiguous broad components in the \hb\ line ($\rm \Delta BIC >100$). The third source, \ruby966323, is fainter and at higher redshift, and has only a marginal detection of a broad component ($\rm \Delta BIC = 1.94$; broad line flux detected at 3.5$\sigma$, $F_{\rm H\beta, broad} = 2.1\pm0.6 \times 10^{-18}\,{\rm erg\,s^{-1}\,cm^{-2}}$). For this source we also perform an independent fit of the Prism spectrum, finding similarly weak evidence for a two-component model and posteriors that are consistent with the fit to the G395M spectrum.

Finally, we fit the Balmer \ha\ emission line, which is available in the G395M spectrum for \ruby49140 and only in the Prism spectrum for \ruby55604 (for \ruby966323 \ha\ is redshifted out of the wavelength range accessible with NIRSpec). \ha\ is modeled with a narrow and broad component, and also fit simultaneously with the \nii$_{\lambda\lambda6549,6585}$ doublet. We assume that \nii\ is narrow and set the width to be equal to that of the narrow \ha\ line, and fix the flux ratio of the doublet to 1:2.94. Moreover, informed by the fits to \hb\ and \oiii, we constrain the dispersion of the narrow lines to $<150\,\kms$. For \ruby49140 we also find a blue-shifted absorption feature in the \ha\ line, which is fit in the same manner as described previously for the \hb\ line of \ruby55604. We note that these fits are only used to obtain an estimate of the broad line flux in order to calculate the broad \ha\ equivalent width (EW) in Section~\ref{sec:results_elines}.

\subsection{Single-epoch Black Hole Mass Estimates\label{subsec:bh_mass}}

Reverberation mapping is a method employed to determine the size of the broad-line region by analyzing the time delay between variations in the AGN continuum and corresponding changes in the broad permitted lines (e.g., \citealt{Blandford1982}). It has enabled the establishment of empirical correlations between the size, line luminosities, and widths in the nearby universe (e.g., \citealt{Kaspi2000,Landt2013}). These correlations facilitate the estimation of black hole masses from single-epoch observations.

Following \citet{Assef2011}, we estimate the black hole mass based on \hb\ and luminosity at rest 5100\,\AA\ as
 \begin{multline}
 	\log (M_{\rm BH}/M_{\odot}) = 0.895 + \\ 0.529 \log \left( \frac{L_{5100}}{10^{44}~{\rm erg~s^{-1}}} \right) + 2.0 \log  \left(  \frac{ {\rm FWHM_{H\beta}} }{{\rm km~s^{-1}}}  \right) .
 \end{multline}
\hb\ is used since \ha\ is not available for all sources and suffers from uncertainty due the blending of the \nii\ doublet with the broad component of \ha. $L_{5100}$ is calculated from the intrinsic AGN spectrum inferred from the SED models. 

We adopt this relation to better illustrate the uncertain in the black hole masses, as the different SED models predict a wide range of AGN luminosities.
All line luminosities here are dereddened using the dust attenuation inferred from SED fitting. While subject to systematic uncertainties, this represents our best estimate of the dust content in the absence of a Balmer decrement.
We note, however, that additional systematic uncertainties are likely introduced by the application of these methods at higher redshifts and in different physical conditions than where they are calibrated (e.g., \citealt{Yue2024}).

\section{Spectral Energy Distribution Modeling\label{sec:prosp}}

The clear detection of the Balmer breaks means that the rest-frame 3500\,\AA\ wavelength range is dominated by starlight. 
However, Balmer breaks are indicative only that evolved stellar populations exist and are relatively bright---they can appear even with ongoing star formation. Therefore, unambiguous Balmer breaks leave some room for interpretation when inferring formation histories.
The mixture of stellar (as suggested by the Balmer break) and AGN continuum emission (as suggested by the broad emission lines) motivate the need for detailed spectrophotometric modeling which includes contributions from both components. 

Given that decomposing galaxy/AGN light is a known challenge, we consider three models to capture the systematic uncertainties in the inferred properties and formation histories. Below we describe the free and fixed parameters of the SED model, and then introduce three different priors which lead to three different interpretations of the observed light.

\renewcommand\thefigure{3-a}
\begin{figure*} 
\gridline{
  \fig{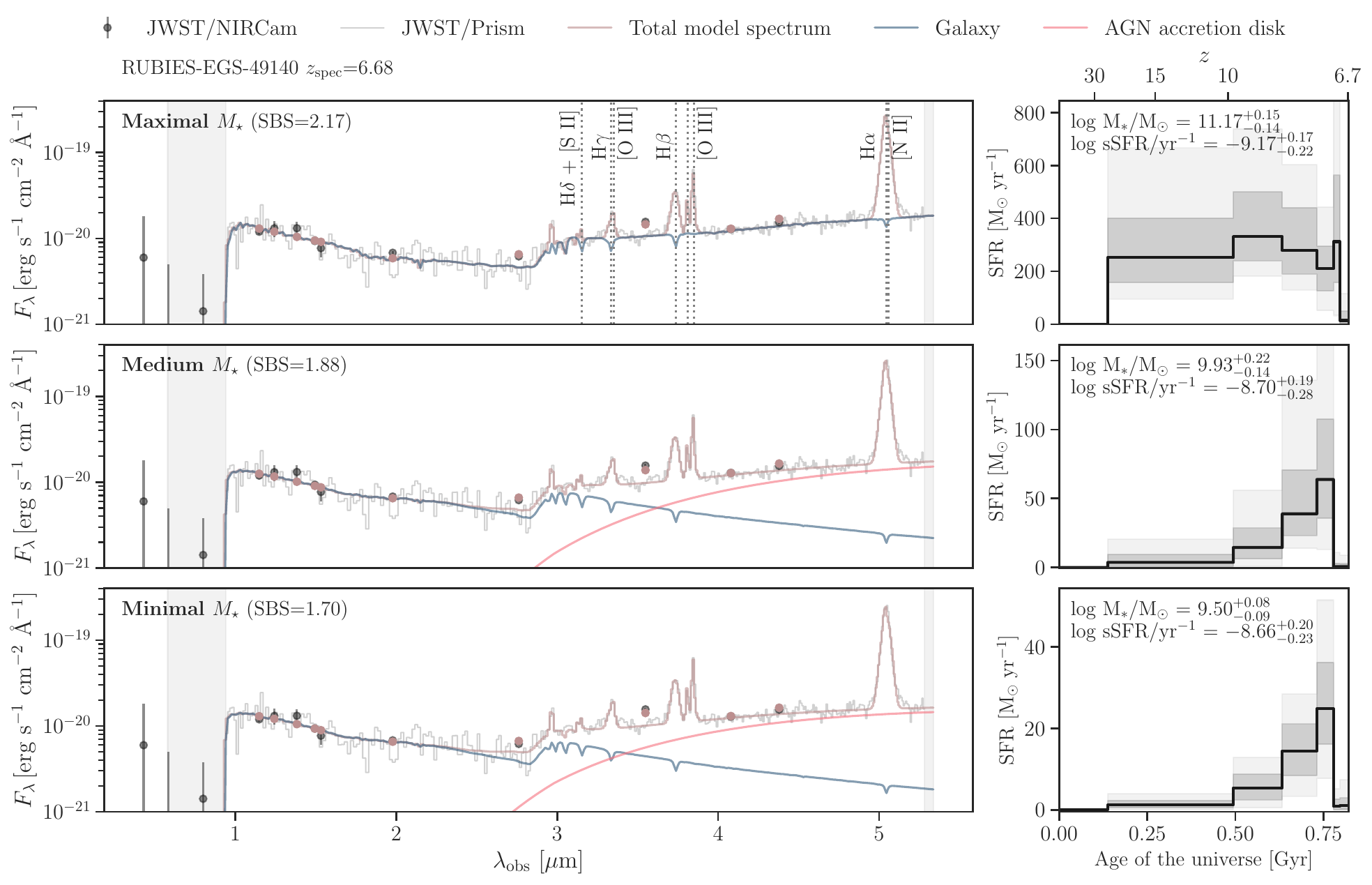}{0.95\textwidth}{}
}
\caption{Spectrophotometric modeling for \ruby49140, with models including maximal / medium / minimal stellar contribution shown. The other two objects with detected Balmer breaks are shown in subsequent figures.
(Left panels) The photometric and spectroscopic data are shown in gray.
The best-fit model spectrum, which includes the marginalized emission lines as annotated, is plotted in light brown. The emission at $\sim 3869$\,\AA\ is likely \neiii\,3869, although He\,\textsc{I}\,3889 is also possible. The galaxy and the AGN continuum components are over-plotted in blue and red, respectively. The spectral break strength, SBS, predicted by the galaxy model spectrum is indicated to the upper left corner. The spectral regions that are masked due to low S/N or detector gap are shaded in gray. 
(Right panels) The inferred SFRs are plotted as a function of the age of the universe. The gray and light gray shading indicates 1$\sigma$ and 2$\sigma$ uncertainties, respectively. The post-starburst feature is primarily driven by the Balmer break.}
\label{fig:sed_a}
\end{figure*}

\renewcommand\thefigure{3-b}
\begin{figure*} 
\gridline{
  \fig{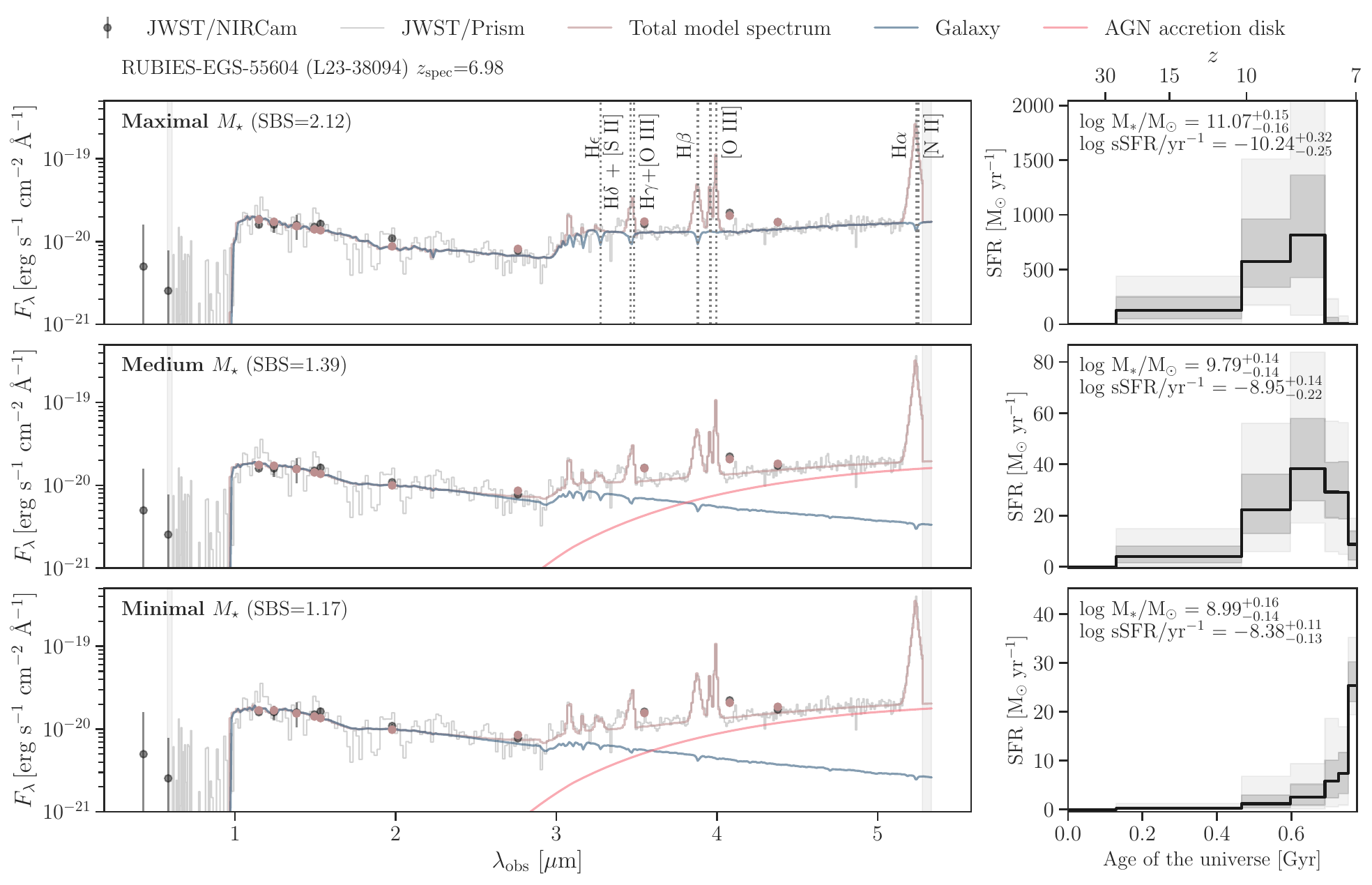}{0.95\textwidth}{}
}
\caption{Spectrophotometric modeling for \ruby55604 (identified as 38094 in L23), with format as described in Figure~\ref{fig:sed_a}.}
\label{fig:sed_b}
\end{figure*}

\renewcommand\thefigure{3-c}
\begin{figure*} 
\gridline{
  \fig{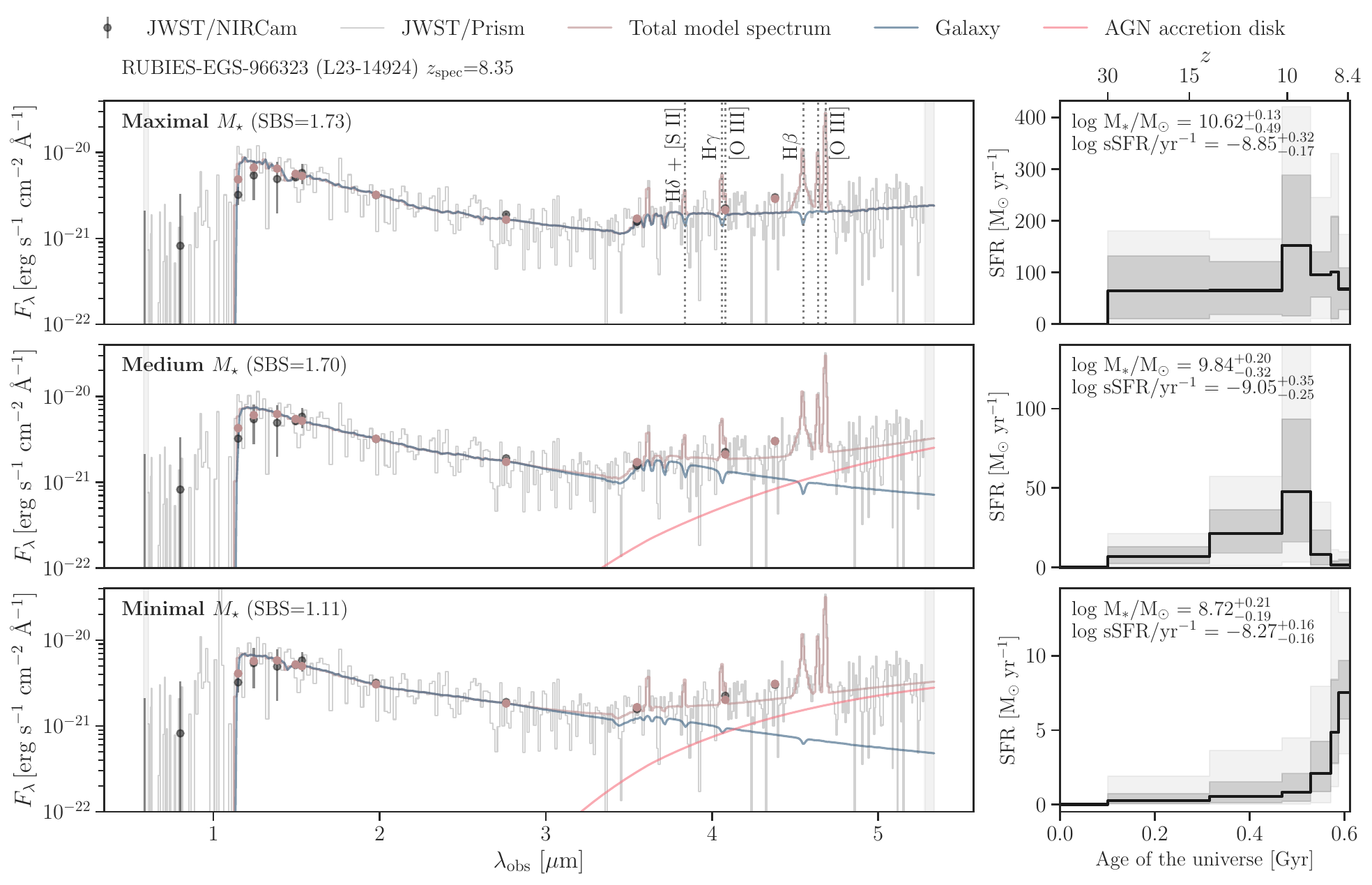}{0.95\textwidth}{}
}
\caption{Spectrophotometric modeling for \ruby966323 (identified as 14924 in L23), with format as described in Figure~\ref{fig:sed_a}.}
\label{fig:sed_c}
\end{figure*}
\setcounter{figure}{3}
\renewcommand\thefigure{\arabic{figure}}

\subsection{Core Model Components}

The available JWST and HST photometric data are jointly fitted with the full NIRSpec/Prism spectrum within the \prospector\ inference framework \citep{Johnson2021}, following \citet{Wang2024:brd}. The setup for the stellar populations is detailed in \citet{Wang2024:sps}. In brief, we adopt the MIST stellar isochrones \citep{Choi2016,Dotter2016} along with the MILES stellar spectral library \citep{Sanchez-Blazquez2006} in FSPS \citep{Conroy2010}. Star formation history (SFH) follows the non-parametric \prospector-$\alpha$ description, characterizing mass formation in 7 logarithmically-spaced time bins \citep{Leja2017}, with a weakly informative prior assumption of a rising SFH from \citet{Wang2023:pbeta}. We also impose a joint prior on stellar mass and stellar metallicity, as introduced in \citet{Leja2019}. This prior is a Gaussian approximation of the relationship measured from the Sloan Digital Sky Survey (SDSS; \citealt{Gallazzi2005}), but the confidence intervals are widened by a factor of 2 to account for potential systematics or redshift evolution. Dust attenuation is described by a \citet{Calzetti2000} curve with a flexible power-law slope \citep{Noll2009}. The fraction of starlight permitted outside the dust screen is allowed to vary, a non-zero fraction of which suggests the presence of blue stars possibly existing outside the dust or having created holes within it. Dust emission is incorporated based on the model by \citet{Draine2007}. 

Model spectra are convolved with the NIRSpec/Prism instrumental resolution curve, tailored for a point source morphology using \texttt{msafit} \citep{deGraaff2023}. We account for wavelength-dependent slit losses by scaling the normalization of the spectrum to match the photometry through a 7th-order polynomial calibration vector applied during the fitting process.

Emission lines are fit using a one-component Gaussian model, and where applicable, a two-component Gaussian model to account for the narrow and broad components. This approach means that the emission lines are not interpreted physically, merely described and included in the modeling of the photometry. To prevent the likelihood from being influenced by residuals from non-Gaussian line kinematics, we enforce a 10\% error floor in the spectroscopic data, which is higher than the conventional 5\% error floor applied to photometry. However, we introduce a multiplicative noise inflation term, with a prior range from 0.5 to 5. Typically, this value is found to be around 1.5, suggesting an additional 50\% inflation of the random noise produces a good fit.
The posteriors are sampled via the dynamic nested sampler \texttt{dynesty} \citep{Speagle2020}.

\subsection{Continuum Contribution from A Dusty Active Galactic Nucleus}

The composite galaxy and AGN model is presented in \citet{Wang2024:brd}, in which we approximate the direct ultraviolet (UV)--optical emission from an AGN accretion disk as piece-wise power laws, with varying normalization.
The slopes are fixed to the best-fit values in \citet{Temple2021}, which are calibrated to the median colors of quasars in SDSS, UKIDSS, and unWISE.
The light from the accretion disk experiences the same dust attenuation as the stars, but is additionally reddened by a separate dust attenuation curve modeled as a power law with varying normalization and shape. In all, the AGN continuum is controlled by 5 free parameters (the AGN-to-galaxy flux ratio, and four dust attenuation parameters, two of which are shared with the galaxy light).

\subsection{Modeling A Range of Possible Stellar and AGN Contribution to the Rest-optical Continua}

The key uncertainty in the interpretation of these objects is the source of the red continuum: is it powered by AGN, stars, or a mixture of both? The implied total stellar masses and formation histories are a strong function of this decomposition. A very red, luminous rest-frame optical stellar solution has a high M/L and an extended formation history, whereas a blue and/or less luminous stellar optical continuum can instead be fit with a flat or rising SFH and relatively little stellar mass. The Balmer break is a key constraint here as this, alongside the resolved sizes in the blue \citep{Baggen2023}, suggest the continuum blueward of rest 4000\,\AA\ is dominated by stars. 

The central question is therefore the origin of the continuum redward of the Balmer break.
We consider three models which bracket the possible range of inferred stellar and AGN contribution to the observed rest-optical continua.

\subsubsection{Maximal Stellar Contribution}

We begin by considering a model which maximizes the stellar contributions. This is achieved by placing a log-normal prior on the pre-dust-attenuated AGN-to-galaxy flux ratio at rest 5500\,\AA, $f_{\rm AGN, 5500A}$, with mean $\mu=-3$, and standard deviation $\sigma=1$. A log-uniform (i.e., flat in log-space) prior is used on the galaxy mass.

This model down-weights the AGN contribution at wavelengths where it is naturally brightest, and effectively leads to stellar masses similar to those from a ``galaxy-only'' fit. While this serves as a useful benchmark, it also implies the broad line EWs, assuming they are driven by AGN, are extremely, perhaps unphysically, high.

\subsubsection{A Mixture of AGN and Stellar Contributions}

We also define a middle ground model, where log-uniform priors are assumed on the total mass formed and $f_{\rm AGN, 5500A}$. 
While it is impossible to write down an agnostic prior, the intent here is to let the data inform the inference process to the maximal extent. This model typically falls between the ``maximum'' and ``minimum'' stellar contributions. However, this does not mean that this model is to be taken as the fiducial model or in some way better motivated than the other choices. The available spectroscopic data are not constraining enough to break the degeneracies among the stellar populations, black hole properties, and dust attenuation.

\subsubsection{Minimal Stellar Contribution}

For the minimal stellar model, we impose a prior on the galaxy mass, with the probability $P(M) \propto M^\alpha$. The slope, $\alpha=-1.7$, is taken to be the low-mass slope of the theoretical stellar mass function at $z=7$ \citep{Tacchella2018}. 
This prior is not intended to exactly replicate the mass function and serves its purpose of producing a low-mass solution even if the true mass function has a different low-mass slope.

A more intuitive prior down-weighting the galaxy contribution would be on the fractional light contribution instead of directly on the stellar mass, as is done with the maximal stellar model. However, the complex translation from light to mass means that, with a prior on the light only, the dust and stellar M/L can adjust to keep the stellar masses similar; a direct prior on the stellar mass avoids this degeneracy.

\section{Results\label{sec:res}}

\begin{deluxetable*}{lllllll}
\tabletypesize{\footnotesize}
\tablecaption{Inferred AGN and Host Galaxy Properties, and Emission Line Kinematics\label{tab:sps}}
\tablehead{\colhead{} &
\colhead{RUBIES-EGS-49140} &  \colhead{RUBIES-EGS-55604} &  \colhead{RUBIES-EGS-966323}
}
\startdata
ID in L23 & -- & 38094 & 14924 \\
RA [deg] &  214.892248 &               214.983026 &               214.876149               \\
DEC [deg] &  52.877410 &                52.956001 &                52.880831                \\
$z_{\rm phot}$ in L23 & -- & $7.48_{-0.04}^{+0.04}$ & $8.87_{-0.09}^{+0.13}$ \\
$z_{\rm spec}$ &  $6.68351_{-0.00009}^{+0.00011}$ &  $6.98173_{-0.00012}^{+0.00013}$ &                     $8.35304_{-0.00015}^{+0.00015}$  \\
\hline
$\log M_\star/\msun$ (max $M_\star$) &  $11.17^{+0.15}_{-0.14}$ &    $11.07^{+0.15}_{-0.16}$ &  $10.62^{+0.13}_{-0.49}$  \\
$\log M_\star/\msun$ (med $M_\star$) &  $9.93^{+0.22}_{-0.14}$ &     $9.79^{+0.14}_{-0.14}$ &   $9.84^{+0.20}_{-0.32}$   \\
$\log M_\star/\msun$ (min $M_\star$) &  $9.50^{+0.08}_{-0.09}$ &     $8.99^{+0.16}_{-0.14}$ &   $8.72^{+0.21}_{-0.19}$   \\
\hline
$\log Z/\zsun$ &                        $-1.22^{+0.20}_{-0.22}$ &    $-0.50^{+0.28}_{-0.32}$ &  $-1.21^{+0.36}_{-0.41}$  \\
Mass-weighted age [Gyr] &               $0.21^{+0.06}_{-0.05}$ &     $0.22^{+0.04}_{-0.04}$ &   $0.22^{+0.03}_{-0.05}$   \\
${\rm SFR_{100}~[\msun\,{yr^{-1}}}]$ &  $30.69^{+21.15}_{-13.37}$ &  $16.80^{+4.69}_{-4.36}$ &  $11.06^{+7.24}_{-5.57}$  \\
$\log {\rm sSFR_{ 100}/{yr^{-1}}}$ &    $-8.70^{+0.19}_{-0.28}$ &    $-8.95^{+0.14}_{-0.22}$ &  $-9.05^{+0.35}_{-0.25}$  \\
$n_{\rm dust,2}$ &                      $-0.21^{+0.22}_{-0.23}$ &    $-0.63^{+0.22}_{-0.22}$ &  $-0.42^{+0.21}_{-0.20}$  \\
$\hat\tau_{\rm dust,2}$ &               $0.63^{+0.41}_{-0.24}$ &     $0.27^{+0.13}_{-0.08}$ &   $3.21^{+0.59}_{-0.78}$   \\
$n_{\rm dust,4}$ &                      $-1.69^{+0.14}_{-0.08}$ &    $-1.68^{+0.11}_{-0.08}$ &  $-1.38^{+0.20}_{-0.16}$  \\
$\hat\tau_{\rm dust,4}$ &               $3.17^{+0.39}_{-0.33}$ &     $3.52^{+0.31}_{-0.36}$ &   $3.01^{+0.55}_{-0.73}$   \\
$f_{\rm agn, 7500A}$ &                  $5.97^{+1.56}_{-1.64}$ &     $4.20^{+1.60}_{-1.10}$ &   $14.59^{+6.91}_{-6.21}$  \\
\hline
$\sigma$ (\hb, broad) [$\kms$] & $1402_{-68}^{+79}$ & $1527_{-106}^{+115}$ &  $1369^{+315}_{-280}$\tablenotemark{\footnotesize{$\dagger$}} \\
$\sigma$ (\oiii, narrow) [$\kms$] & $68_{-30}^{+17}$ & $48_{-27}^{+20}$\tablenotemark{\footnotesize{$\ast$}} &  $69^{+15}_{-27}$ \\
\enddata
\tablenotetext{$$^\ast$$}{Narrow component of the two-component fit to the \oiii\ line.}
\tablenotetext{$$^\dagger$$}{Ambiguous broad component.}
\tablecomments{For the inferred model parameters excluding the stellar mass, only those from the medium-stellar model are included.}
\end{deluxetable*}

Here we describe the properties resulting from the individual fits to the three objects with Balmer breaks detected at $\zspec = 6.7 - 8.4$. As alluded to in Section~\ref{sec:prosp}, even in the presence of an unambiguous Balmer break, there can be a wide range of inferred star formation histories. The Balmer break strengths, as measured by the spectral break strength defined in Section~\ref{sec:balmer_break}, vary from 2.2 to 1.7 (Figure~\ref{fig:sed_a}), 2.1 to 1.2 (Figure~\ref{fig:sed_b}), and 1.7 to 1.1 (Figure~\ref{fig:sed_c}), by altering the AGN contribution from $\sim 0$ to $\sim100$\% for \ruby49140, 55604, and 966323, respectively.

The range of spectral break strengths also suggests that the adopted sets of SED models bracket the uncertainty in continuum decomposition. As summarized in Table~\ref{tab:sps}, the medium stellar model typically leads to $M_\star\sim10^{9.8}~\msun$. If the red optical light is instead driven by an older dusty stellar population, M/L can rise dramatically, producing masses up to $M_\star\sim10^{11}~\msun$. In the other direction, a prior that minimizes the inferred stellar masses can still reproduce the light while lowering the inferred masses by up to 1 dex.

In this section we lay out the physical properties and the implications of the formation histories inferred by the three physical models, including both observational and evolutionary considerations. The aim here is to provide an unbiased, comprehensive view into the very different possible physical interpretations of these objects---all of which result in statistically acceptable fits to the observed spectra.

\subsection{Apparent Wavelength-dependent Contribution of AGN and Stellar Light\label{sec:results_elines}}

The models with minimal and medium stellar contribution infer that the AGN accretion disk dominates the rest-optical, while starlight dominates the rest-UV and also the Balmer break (by necessity). The AGN emission is heavily attenuated by dust, resulting in the observed red color. The transition from stellar- to AGN-dominated light can in principle be corroborated by the wavelength-dependent morphology---an unresolved morphology is more indicative of AGN-dominated light. The presence of broad emission lines is both a suggestion of AGN activity and, if these broad lines are AGN-powered as opposed to e.g., a stellar-driven outflow, they can put a soft lower limit on the AGN continuum contribution. This is because the AGN-powered broad H$\alpha$ EWs are typically observed to be $\lesssim 1000$\,\AA\ (more discussion of this point in Section~\ref{sec:concl}); decreasing the AGN contribution to the continuum underneath these broad lines requires correspondingly higher AGN EWs, potentially outside of the previously observed range, which would require different and new AGN physics.

\begin{figure} 
\gridline{
  \fig{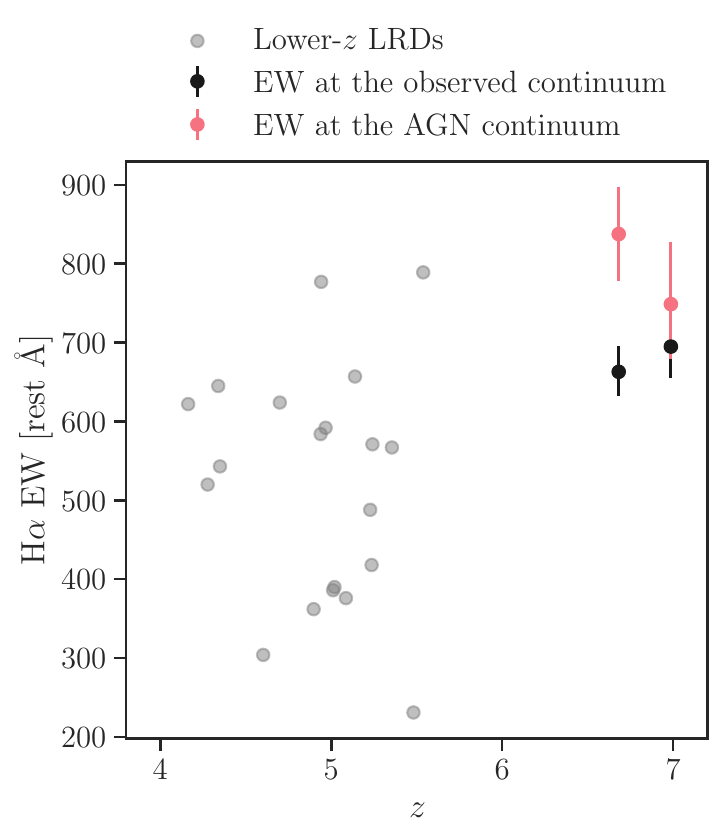}{0.45\textwidth}{}
}
\caption{Rest-frame equivalent widths of \ha. EWs of the broad component of \ha\ measured for the sample of this paper at the total continuum and AGN continuum (assuming the minimal stellar model) are represented in black and red, respectively. For reference, the EWs of a grism-selected LRD sample are plotted in gray \citep{Matthee2023}. Assuming the broad lines trace the AGN, then the implied EW at the AGN continuum in the case of starlight dominating the rest optical would be orders of magnitudes higher than the observed EW. Conversely, an AGN-dominated interpretation puts the EWs of our sample in a similar range as the grism-selected LRD sample.
}
\label{fig:ew}
\end{figure}

We find unambiguous broad \hb\ lines with FWHM $>2500~\kms$ in \ruby49140 and 55604, which would be unusual for stellar-driven outflows (see \citealt{Veilleux2005} for an overview, or e.g., \citealt{Heckman2015,Wang2020} for local starbursts). Stellar-driven outflows would typically also be seen in the forbidden lines, which are narrow in these objects---though it is possible for the outflowing gas to be sufficiently dense that it is not luminous in the forbidden lines. A broad component is marginally detected in \ruby966323 (\S\,\ref{sec:lines}). 
Additionally, we compare the observed \ha\ EW to previous samples of LRDs. \ruby49140 and 55604 show large observed EWs of $>600$\,\AA.
We re-estimate the EW using the AGN continuum from the SED fits. Assuming the broad \ha\ is originated from an AGN, the intrinsic EW must be larger in the models where considerable galaxy light contributes to the continuum near \ha. 
As seen from Figure~\ref{fig:ew}, the increased EWs at the AGN continuum inferred from the medium stellar model are still roughly consistent with the distributions measured from a grism-selected LRD sample at lower redshifts \citep{Matthee2023}, although at the higher end. 

We can also predict \ha\ fluxes from the galaxy, given the inferred stellar population parameters, which provide additional leverage on interpreting the continuum emission.
For all three SED models, the predicted \ha\ fluxes are at least 1 order of magnitude lower than the observed flux. This means that the SFRs in all models are much lower than the observed total \ha\ flux, i.e., none of the stellar components in these models can be ruled out by requiring star formation which produces more \ha\ flux than is observed. Conversely, the observed large \ha\ EW suggests an abundance of ionizing photons, the evidence of which is not obvious from the spectra in the UV--optical.

Furthermore, we observe an emission line at $\sim 3869$\,\AA\ in all three objects. One possibility is He\,\textsc{I}\,3889, given that strong He\,\textsc{I} emission line has been reported in spectra of LRDs \citep{Wang2024:brd}, and AGNs more generally \citep{Riffel2006,Landt2008}. More likely it corresponds to \neiii, in which case it would suggest high ionization state, particularly considering the weak (non-detected) \oii\,3727 emission. A strong \neiii/\oii\ ratio can also be indicative of AGN activity \citep{Backhaus2024}.

\begin{figure*} 
\gridline{
  \fig{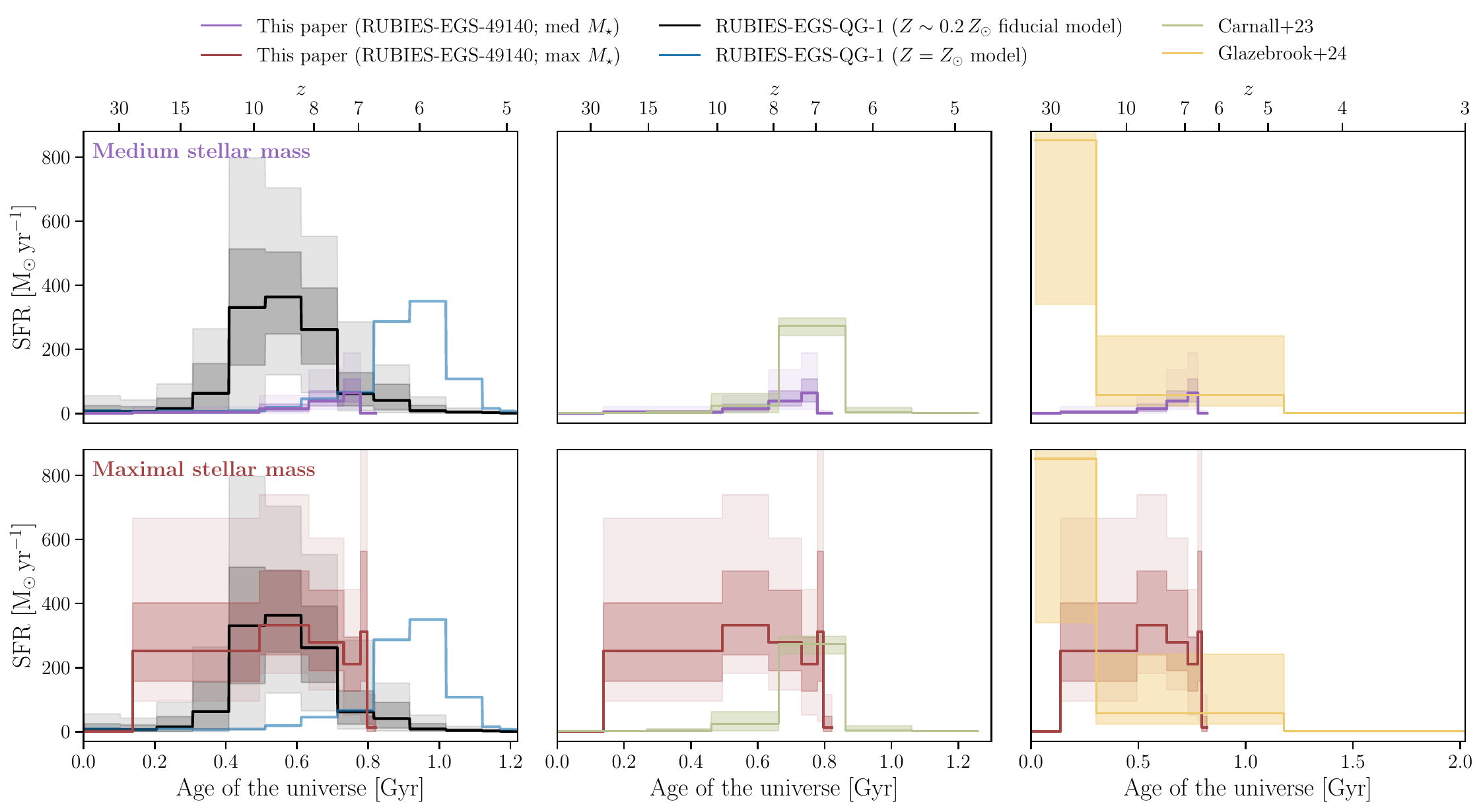}{0.95\textwidth}{}
}
\caption{Formation history inferred for \ruby49140, compared to those inferred for maximally old quiescent galaxies at $z\sim 3 - 5$.
From left to right, we show the SFHs of RUBIES-EGS-QG-1 (at $z=4.9$ with $10^{11}$ M$_{\odot}$; \citealt{deGraaff2024}), GS-9209 (at $z=4.7$ with $4 \times 10^{10}$ M$_{\odot}$; \citealt{Carnall2023}), and ZF-UDS-7329 (at $z=3.2$ with $10^{11}$ M$_{\odot}$; \citealt{Glazebrook2023}). For the the $z=4.9$ galaxy, we include the fiducial low-metallicity solution in black, with $1\sigma$ and $2 \sigma$ uncertainties shaded in grey, and also an alternative solution assuming a solar metallicity in blue. For the other two quiescent galaxies, we only show the $1\sigma$ uncertainty.
The top panels show the formation history from our medium-stellar model while the bottom panels show the formation history from our maximal-stellar model. A stellar interpretation of the light in \ruby49140 produces greater consistency with the massive quiescent galaxies, while the medium model fails to predict sufficient stellar mass to connect \ruby49140 to these massive old populations at lower redshift.}
\label{fig:sfh}
\end{figure*}

\begin{figure*} 
\gridline{
  \fig{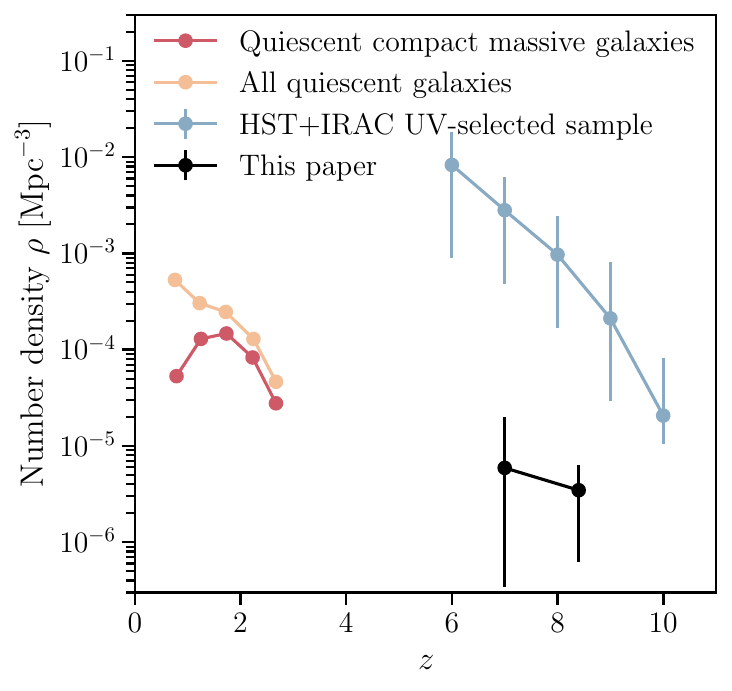}{0.32\textwidth}{(a)}
  \fig{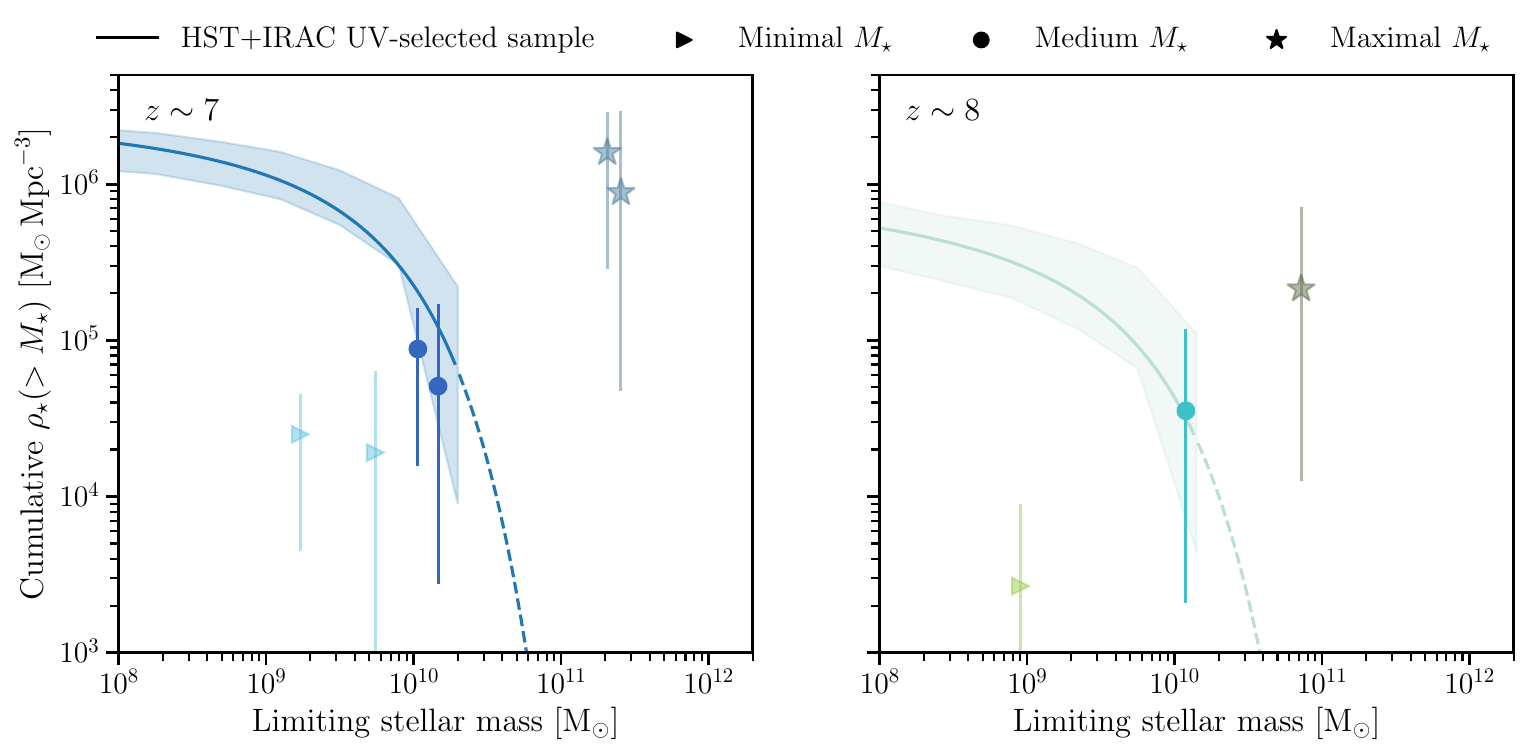}{0.67\textwidth}{(b)}
}
\caption{Implied number density and  mass density.
(a) Number density of different samples. The red and orange curves are taken from \citet{vanDokkum2015}. The blue curve is the total number density for UV-selected sample, obtained by integrating the Schechter fits down to $10^8~\msun$ \citep{Stefanon2021}. Number density for the sample of this paper assuming the EGS area for the volume, in two redshift bins ($5.5 < z \leq 7.0$, and $7.0 < z \leq 8.5$), are shown in black; uncertainties reflect Poisson statistics and cosmic variance \citep{Gehrels1986}. This illustrates, preliminarily, that the number density of the Balmer break sample are comfortably below that of the $z=3$ compact quiescent cores.
(b) Cumulative cosmic stellar mass density, $\rho_\star$. The curves are derived from the same Schechter fits, with the extrapolated regions indicated in dash. $\rho_\star$ from this work are reported for the three SED models.
Uncertainties again reflect Poisson statistics and cosmic variance.
To facilitate a direct comparison to the \citet{Stefanon2021} stellar mass functions, we adjust all our Chabrier stellar masses by $+0.24$~dex to Salpeter masses \citep{Salpeter1955}.
While the number density of the Balmer break sample is low, the mass density in these objects at least comprise $\sim1\%$ of the cosmic mass budget (albeit with large uncertainties; $\sim 0.1\%$ within $1\sigma$). With the medium-stellar model, their contribution is roughly equal to the mass density in all UV-selected objects combined.
Note that we do not attempt to account for selection effects, so all the reported densities for our sample should be taken as lower limits (see \S\,\ref{subsec:dis:rho} for details).
}
\label{fig:rho}
\end{figure*}

\begin{figure*} 
\gridline{
  \fig{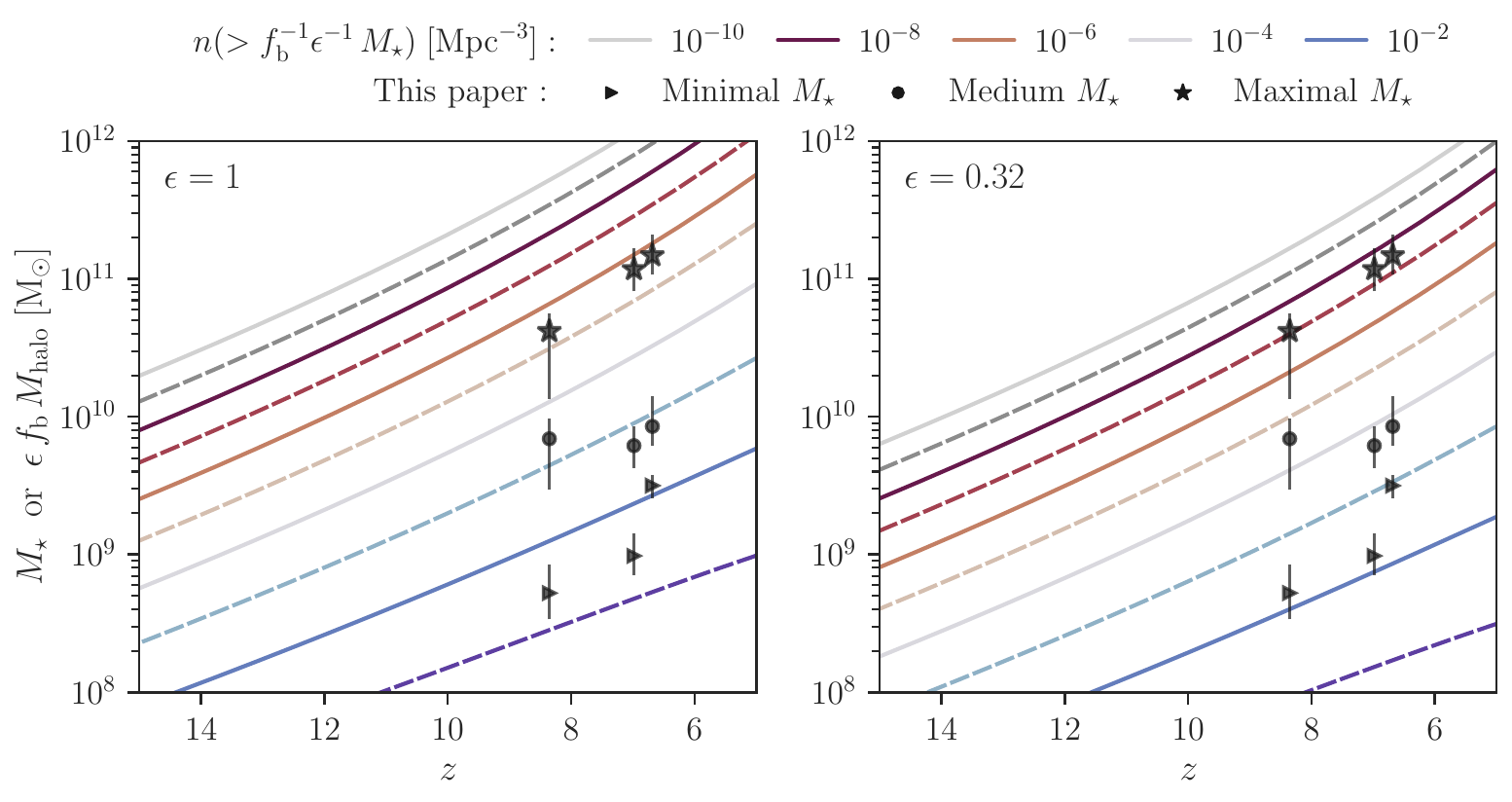}{0.49\textwidth}{(a)}
  \fig{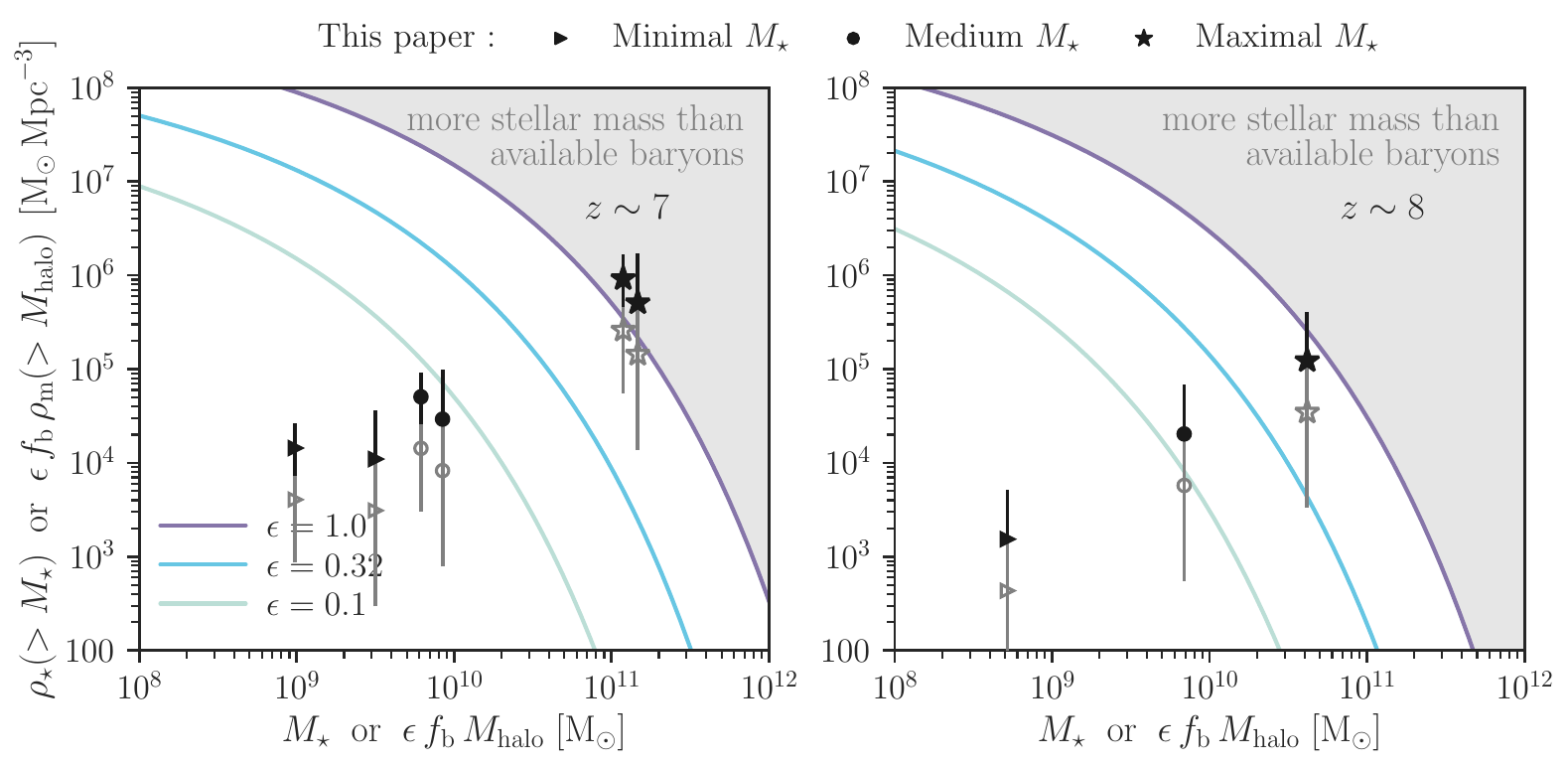}{0.49\textwidth}{(b)}
}
\caption{Comparison to theoretical limits.
(a) Limits on the abundance of galaxies as a function of redshift. Curves illustrate $M_\star$ as functions of $z$ at fixed cumulative halo abundance, assuming the physically maximal $\epsilon = 1.0$ to the left, and the more likely case of $\epsilon = 0.32$ to the right. These plots suggest that it would be rare to find galaxies as massive as the maximal stellar mass case in these redshifts.
(b) Stellar mass density limits. Similar to Figure~\ref{fig:rho}-b, but we now include the theoretical co-moving stellar mass density contained within galaxies more massive than $M_\star$ in the two redshift bins for three values of $\epsilon$. The maximal stellar mass case implies an an unrealistic limit that all available baryons in the haloes are converted into stars.
For completeness, the naive estimates of the lower limits on the number density, based on the CEERS+UDS volume (instead of considering the CEERS field alone as in the fiducial case), are included here as unfilled gray markers.
}
\label{fig:mhalo}
\end{figure*}

\begin{figure} 
\gridline{
  \fig{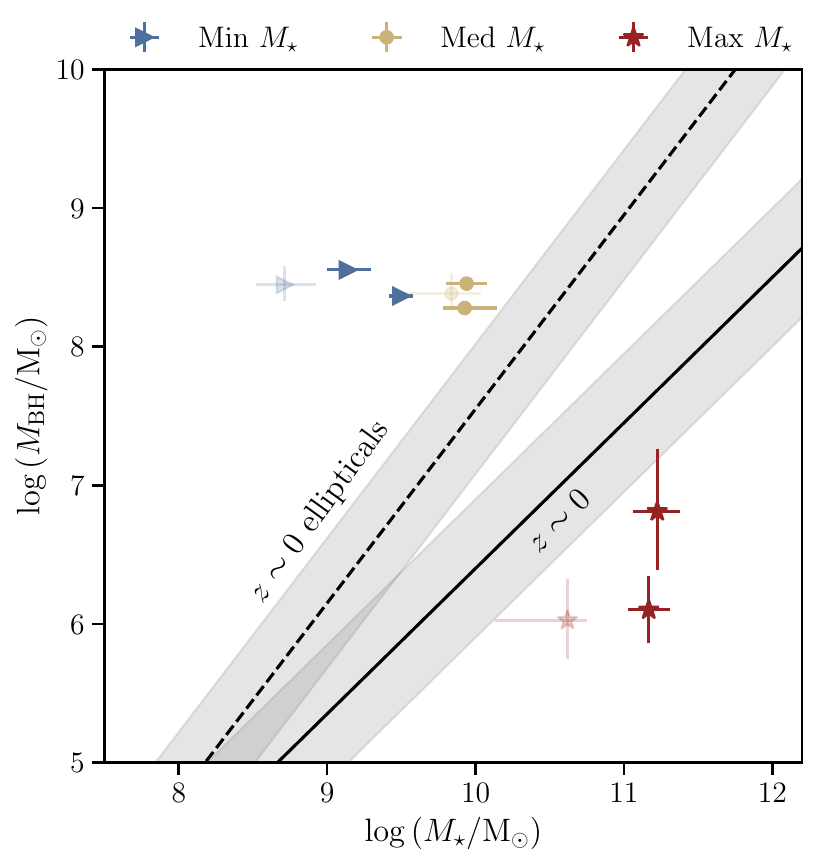}{0.45\textwidth}{}
}
\caption{Stellar--black hole mass scaling relationship. The $z\sim0$ (solid line) and $z\sim0$ ellipticals (dashed line) relations correspond to the black hole mass and host galaxy total stellar mass scaling, and non-active ellipticals with dynamical black hole mass measurements, respectively \citep{Reines2015}. The gray shading indicates the intrinsic scatter. Over-plotted are the black hole masses of our sample estimated from $L_{5100}$ and the broad \hb\ lines, and their corresponding stellar mass inferred from three SED models. The semi-transparent points indicate \ruby966323 for which the presence of a broad line is ambiguous. The black hole masses are uncertain as the inferred AGN luminosity from the SED models vary.
}
\label{fig:mbh}
\end{figure}

Finally, we cross-check the morphologies of \ruby55604 and 966323 as studied in \citet{Baggen2023}. \ruby55604 (L23-38094) is among the most compact in L23, which is unresolved in the LW filters ($<0.22$~kpc), but clearly exhibits two components in F115W and F150W. \ruby966323 is also unresolved in F444W, and likely resolved in F200W, although the F200W data is too faint to have a robust size estimate. Preliminary analysis on \ruby49140 reveals the same trend. 
A thorough study on the morphologies will be presented in J. Baggen et al. in prep.
These findings again indicate the presence of a transition from stellar to AGN light near the spectral break, thus favoring the minimum/medium stellar models.

\subsection{Inferred Formation Histories}

The inferred SFHs based on the three SED models are shown in Figure~\ref{fig:sed_a}--\ref{fig:sed_c}. All models have significant SFHs extending over hundreds of millions of years, necessary to produce the evolved stellar populations responsible for the Balmer breaks. However, there is some variation in the timescales of star formation, and significant variation in the amplitudes.
The medium and maximal-stellar models require significant star formation at $z=10$ and a recent decline in the SFR. The decline may be representative of a very early termination in star formation necessary to produce the very old galaxies observed at lower redshifts \citep{Carnall2023,Glazebrook2023} or a more temporary mini-quenching event \citep{Looser2023}; however, we note that at least two of these Balmer breaks are substantially stronger than the \citet{Looser2023} break, implying a correspondingly longer period of quiescence ($\gtrsim 100$~Myr) than a mini-quenching event.

Conversely, the minimal-stellar models decrease the strength of the stellar Balmer breaks by assuming a very red AGN continuum underneath, and in this way can infer purely rising SFHs for \ruby55604 and 966323. The signature of a recent decline in the SFH is still preserved in the brightest source, \ruby49140, which also has the strongest Balmer break among the three. This is an expected behavior---the less prominent Balmer breaks may be fit with a weaker break and steeper AGN continuum emission as the prior drives to minimize the stellar mass. However, it is worth emphasizing that this scenario hinges on a peculiar AGN continuum shape, which coincidentally aligns perfectly with that of the host galaxy to produce the observed spectral break. Alternatively, if not all the UV light is from stars, (e.g., scattered light from the AGN; \citealt{Greene2023}), this would remove an important constraint pushing the inferred stellar ages younger, and perhaps relax the stringent requirement on the galaxy and the AGN conspiring to create a spectral break by allowing an older stellar population.

One potential concern may arise from the age--metallicity degeneracy. We note that the data do not constrain the metallicity, as the posteriors roughly follow the prior distributions. This effectively means that we marginalize over metallicity, i.e., the uncertainty in metallicity is propagated into the rest of the inferred stellar population parameters.

Interestingly, the maximal-stellar interpretation can be more easily connected with several massive, very old quiescent galaxies recently discovered with JWST at $z=3-5$ \citep{Carnall2023,Glazebrook2023,deGraaff2024}, as illustrated in Figure~\ref{fig:sfh}. Such a link may serve as an argument in favor of the maximal-stellar model.

We note, though, the above connection is still hard to establish given the low number densities of these massive objects (a lower limit of $\sim10^{-5}$~Mpc$^{-3}$ for the three objects in this work, and e.g., $\sim 5 \times 10^{-6}$~Mpc$^{-3}$ for \ruby QG-1 estimated in  \citealt{deGraaff2024}). Nevertheless, the most straightforward prediction stemming from the discoveries of these quiescent galaxies at $z=3-5$ is the presence of Balmer breaks at $z=7-8$, suggesting very early and/or rapid formation. While the number density may be low, they are indicative of a class of objects with early and rapid formation.

\subsection{Contribution to the Cosmic Stellar Mass Density\label{subsec:dis:rho}}

One of the key results from L23 is that the high stellar masses in these objects suggest they dominate not just the high-mass end of the pre-JWST UV-selected galaxy stellar mass function \citep{Stefanon2021}, but indeed the entire stellar mass budget at this cosmic epoch. We revisit this point by estimating the cumulative stellar mass density using the stellar masses inferred in this work. Specifically, we bin the three objects into two redshift ranges ($5.5 < z \leq 7.0$, and $7.0 < z \leq 8.5$), and sum the mass in each bin. The cosmic volume is estimated by integrating between the redshift limits over the areal coverage of CEERS (88.1 arcmin$^2$; \citealt{Finkelstein2023}). As in L23, we only consider the Poisson uncertainty and cosmic variance, and neglect corrections for incompleteness, given that any correction for incompleteness would increase the inferred stellar mass densities.

We note that there is some evidence indicating that EGS exhibits an over-density of red objects. So far, nothing similar to the sample of this paper, in terms of high-$z$ Balmer breaks in the available spectra, and more broadly, of objects as bright and red in our parent photometric catalogs, has been identified in other RUBIES fields (i.e., UDS, which has an area of 224 arcmin$^2$; \citealt{Weibel2024}). A naive estimate of the number density might include this larger empty volume. However, as will become evident subsequently, this factor of $\sim 3$ decrease does not affect any of the main conclusions. A subtle point, though, is that for objects displaying very red colors in broadband photometry, their spectra may not necessarily be similar. Defining volumes using photometry might therefore be suboptimal, but it perhaps can serve as a reasonable approximation here given the strong correlation between stellar masses and luminosity/flux.

Figure~\ref{fig:rho} shows the cumulative stellar mass density in these objects compared to rest UV-selected objects. Importantly, rest-UV selection would fail to select any of the objects in this paper, which are extremely red: the total mass density is thus inferred to be the sum of the two (independent) curves. Although stellar mass functions incorporating the JWST observations have been estimated (e.g., \citealt{Harvey2024,Weibel2024}), the unclear physical nature of the reddest objects makes it especially difficult to both estimate and interpret the high-mass end of the high-redshift stellar mass functions. We thus compare only to the pre-JWST UV-selected mass functions, as a test of what a rest-UV selection function may have missed.

Assuming the medium-stellar model, our analysis suggests that at $z\sim7-8$, this small population contains $20-50\%$ of the stellar mass density of the entire rest-UV selected galaxy population, implying that $\sim20-50\%$ of star formation at $z>7$ occurs in the progenitors of these optically red objects and that they would dominate the high-mass end of the mass function.
Adapting the minimal-stellar model, these objects comprise a much smaller $\sim1\%$ of the cosmic mass budget and live in much more typical galaxies.
Given that the Balmer break objects may have slowing or declining star formation histories, whereas UV-selected galaxies typically have rising formation histories by nature, this fraction is likely to be even higher at higher redshifts.

For reference, the number density of our sample is more than an order of magnitude lower than the total number density of the rest-UV selected sample, obtained by integrating the Schechter fits down to a stellar mass limit of $10^8~\msun$ \citep{Stefanon2021}. This reinforces that the Balmer break objects may host disproportionately high levels of past star formation compared to the general galaxy population.

\subsection{Comparison to Theoretical Limits\label{subsec:dis:halo}}

We now put the above stellar mass density into context with theoretical predictions from the standard $\Lambda$CDM cosmology.
Given the suite of cosmological parameters, the matter power spectrum describing the density contrast of the universe on large scales can be specified. As gravitational collapse becomes non-linear on smaller scales which are relevant for dark matter haloes that host galaxies, higher-order statistics becomes necessary. However, under the assumption that the initial density fluctuations are Gaussian and small, reasonable analytic predictions of the halo distribution can be made \citep{Press1974}. 

Following \citet{Boylan-Kolchin2023}, we adopt the \citet{Sheth1999} halo mass function, which is an extension of the \citet{Press1974} formalism by assuming ellipsoidal collapse instead of the more simplified spherical collapse. The co-moving number density of haloes above a given halo mass threshold, $M_{\rm halo}$, is estimated as 
\begin{equation}
	n (>M_{\rm halo}, z) = \int_{M_{\rm halo}}^{\infty} {\rm d}M \frac{{\rm d}n(M,z)}{{\rm d}M },
\end{equation}
where ${\rm d}n(M, z)/{\rm{d}}M$ is the number of dark matter haloes of mass $M$ per unit mass per unit co-moving volume at redshift $z$ (i.e., the halo mass function).
Then the co-moving mass density in haloes more massive than $M_{\rm halo}$ can be computed straightforwardly as
\begin{equation}
	\rho_m (>M_{\rm halo}, z) = \int_{M_{\rm halo}}^{\infty} {\rm d}M M \frac{{\rm d}n(M,z)}{{\rm d}M }.
\end{equation}

From the above, we can obtain the corresponding statistics of galaxies by assuming
\begin{equation}
	M_\star = \epsilon f_{\rm b} M_{\rm halo},
\end{equation}
where $\epsilon$ is the efficiency of conversion of baryons into stars, and $f_{\rm b}$ is the cosmic baryon fraction. $\epsilon=1$ sets the stringent upper limit on the stellar content that a halo can have. We also consider two more likely values of $\epsilon=0.1, 0.32$ (e.g., \citealt{Behroozi2013,Moster2013}).

As illustrated in Figure~\ref{fig:mhalo}, only in the limiting case of $\epsilon = 1.0$ does the expected number density of galaxies with the maximal stellar mass approximately align with the $\Lambda$CDM prediction, corresponding to haloes with cumulative co-moving number densities $\lesssim 10^{-5.4} ~{\rm Mpc}^{-3}$).
For $\epsilon = 0.32 ~ (0.10)$, the implied number density is $\lesssim 10^{-7.1} (10^{-9.5}) ~{\rm Mpc}^{-3}$, suggesting that the maximal stellar mass model results in objects that are unexpectedly massive. Furthermore, similar to findings in \citet{Boylan-Kolchin2023}, the implied star formation efficiency from our maximal stellar mass estimate lies at the extreme end of $\Lambda$CDM expectations. These tensions are alleviated if assuming instead the minimal and medium stellar mass models.
For completeness, we also include the second set of lower number density, based on the CEERS+UDS volume, as unfilled gray markers in Figure~\ref{fig:mhalo}-b.
Additional systematic uncertainties in SED fitting, e.g., the initial mass function which can likewise decrease the tension with the standard model, are presented in \citet{Wang2024:sys}.

\subsection{Dynamical Mass}

While the Balmer lines are very broad, the forbidden lines are relatively narrow: the \oiii\ line widths for the three galaxies are FWHM $\sim$ 160, 113, and 162 $\kms$ (Table~\ref{tab:sps}). 
As gas is viscous, the forbidden lines likely originate in disks, and---assuming those disks trace the gravitational potential of the galaxies---can be used to derive dynamical masses. As discussed in, e.g., \citealt{ForsterSchreiber2014,vanDokkum2015,Price2016}, the conversion from line width to dynamical mass depends on the spatial distribution of the gas, the orientation of the disks, and the orientation of the slit with respect to the rotation axis. While the gas is certainly compact (based on 2D spectra, where \oiii\ lines have about the same point-like morphology as the Balmer lines), the latter two dependencies are both unknown. 

Nevertheless, it is instructive to estimate the dynamical mass. Following \citet{vanDokkum2015}, we have
\begin{equation}
	M_{\rm dyn} = 2 \frac{V_{\rm rot}^2 r_{\rm e}}{G},
\end{equation}
where $r_{\rm e}$ is the spatial scale, taken to be $\sim 100$ pc \citep{Baggen2023}, and $V_{\rm rot}$ is the rotation velocity derived via
\begin{equation}
	V_{\rm rot} = \frac{\sigma_{\rm gas}}{\alpha \sin^{-1}(i)}.
\end{equation}
Assuming $\alpha=0.8$ \citep{Rix1997,Weiner2006}, and an inclination angle, $i$, of 45 degrees, the dynamical masses for \ruby49149, 55604, and 966323 are $\sim 10^{8.6}$, $10^{8.3}$, and $10^{8.6}~\msun$, respectively, all at the low end of the stellar mass estimates, and actually in some tension with the black hole mass estimates alone (discussed in the next section).

Higher masses are possible if the gas has a larger spatial extent than the stars \citep{vanDokkum2015}, though the \oiii\ emission lines appear to have a highly compact morphology in the 2D spectra. Importantly, the LSF of NIRSpec is strongly dependent on morphology (up to a factor of $\approx2$ difference in resolution between a point source and uniformly illuminated slit; \citealt{deGraaff2023}). Therefore, if the source were to have $r_{\rm e} \sim 1$ kpc, then $\sigma_{\rm gas}$ would decrease to $< 30 ~\kms$; that is, by making the source larger, the dynamical mass remains similarly small.
Low disk inclinations can also increase the dynamical mass. 
This is unlikely to be the case by chance, but it could result from selection effects; the red color selection may preferentially select obscured AGNs at particular down-the-barrel (i.e., face-on) orientations. However, given the significant inferred dust attenuation, a down-the-barrel orientation would be in some tension with the standard AGN model.

\subsection{Galaxy--Black Hole Scaling Relationship\label{subsec:bh}}

Figure~\ref{fig:mbh} compares the stellar mass--black hole mass relationship inferred for these objects to the local relationships. In the optically red regime where our sample resides, the M/L--color relationship used for the local relationships \citep{Zibetti2009} agree well with the \prospector\ M/L--color relationship \citep{Li2022}.

Based on our minimal/medium stellar masses, these objects host massive black holes $\sim 10 \times$ above the $z=0$ $\mstarmbh$ relationship \citep{Reines2015}. The minimal stellar mass model is well above the typical scatter in this relationship, while the medium model is within $\sim2\sigma$.
These results align with previous findings on single objects \citep{Kokorev2023,Furtak2023:spec}, quasars \citep{Stone2024}, as well as that a $z\sim5$ AGN sample which is located $\sim 10 - 100 \times$ above the local relationship \citep{Pacucci2023}. At face value, this implies that these black holes are over-massive relative to their stellar components, and that the stellar components must continue to grow over the next 13 Gyr in order to produce the relationship observed today.
However, we caution that the selection bias, where the luminous and massive black holes with emission sufficiently dominating their host galaxy are more readily observed in a flux-limited survey \citep{Lauer2007}, is likely exacerbated by the high-redshift frontier.

Meanwhile, the maximal stellar model predicts an AGN continuum orders of magnitudes lower than the galaxy continuum, and hence implies much lower black hole masses using the $L_{5100}$ relation, although as noted above, it is difficulty to know how to calculate the black hole mass when the SEDs are so abnormal. Alternatively, since this model effectively predicts starlight dominating over all the observed wavelength range, it may be possible that these galaxies do not host AGNs at all, which is reasonable given the non-detection in X-rays \citep{Ananna2024,Yue2024}. This then would perhaps require some yet-to-be-understood mechanism to produce the broad emission lines---for example, broad emission lines from supernovae have been previously mistaken for quasar activity \citep{Filippenko1997,Aretxaga1999,Baldassare2016}, though it remains unclear why SNe would be consistently associated with the other spectral characteristics of these objects, and such interpretation may be in some tension with the lack of variability seen in LRDs \citep{Maiolino2024:xray}.

While it is instructive to put our objects in the context of the $\mstarmbh$ relationship, despite substantial uncertainty in both the black hole and galaxy mass, we note that the $z\gtrsim 7$ objects do not necessarily need to lie on the $z=0$ relationship. \citet{Peng2007} proposed that a large number of mergers lead to a statistical convergence process, and thus the slope of the $\mstarmbh$ relationship always becomes $\sim 1$, regardless of the initial condition.
In any case, the above scenarios all point to very different pictures of the early $\mstarmbh$ relationship and the subsequent evolution required to match the local scaling laws.

\section{Discussion and Conclusions\label{sec:concl}}

Having discussed the implications of the physical properties and the formation histories resulting from each model, we now attempt to tie all the pieces together. 
We begin this section by briefly summarizing the modeling results, and then examine two potential interpretations of the nature of these objects.

\subsection{A Brief Summary of the Implications of Different Stellar Contributions}

The maximal stellar model, if correct, would have a remarkable impact on the first billion years of galaxy evolution. It proposes that the cumulative stellar mass densities in these objects is comparable to that of all the UV-selected objects at these redshifts (Figure~\ref{fig:rho}). Such early and efficient formation would be in tension with the standard assumption on the baryon to stellar conversion efficiency (Figure~\ref{fig:mhalo}; also see \citealt{Boylan-Kolchin2023}). Yet such an interpretation is consistent with the very old stellar populations observed in high-mass galaxies at $z=3-5$ (Figure~\ref{fig:sfh})---these require rapid, early assembly of $\gtrsim10^{10.5}$ M$_{\odot}$ in stellar mass, followed by an early cessation in star formation. The SEDs of their progenitors would look much like these objects---though these objects are $\sim10 \times$ smaller in physical size than these later quiescent galaxies.

In contrast, the models with minimal and medium stellar contributions (which in turn imply larger fractional AGN contributions) are supported by the observations of broad emission lines and compact morphologies at redder wavelengths. They require the black hole continuum to be steeply rising in the rest-optical, such that a rapid transition from a stellar-dominated continuum to a black hole-dominated continuum occurs.
Furthermore, barring line-of-sight or other arguments, the small dynamical mass is in distinct tension with all models except the minimal stellar contribution. Meanwhile, the medium stellar model leads to a $\mstarmbh$ relation in less tension with (or requiring fewer mergers to become consistent with) the local relations (Figure~\ref{fig:mbh}).

\subsection{Progenitors of Massive Quiescent Galaxies, or Low-mass Galaxies with Bright Black Holes?}

It has long been suspected that the cores of the most massive galaxies in the local universe formed in a spectacular early burst of star formation at $z\gtrsim5$, followed by an immediate and permanent quenching. This is underpinned by their high $\alpha$-element abundances and their old (but unresolved) ages (e.g., \citealt{thomas05,Conroy2014}), and from number density analysis of massive, dense galaxy cores at $z=0-3$ \citep{vanDokkum2015}. 

Further evidence has risen to support this interpretation in the JWST era, in the form of ``maximally old'' quiescent galaxies identified at $z=3-5$ with formation redshifts $z>10$ \citep{Glazebrook2023}, $z\sim10$ \citep{deGraaff2024} and $z\sim8$ \citep{Carnall2023}. The higher redshift, that these objects are observed at, allows for better resolution of the formation timescales, due to the younger age of the universe. This puts the formation of at least some of these massive galaxies in the first 500--600 Myr after the Big Bang. These objects have high stellar masses of 0.4--1 $\times 10^{11}~\msun$, and must have quenched shortly after the formation of the bulk of their stellar mass, but likely progenitors have yet to be conclusively identified at this epoch.

The above gap in the observed evolution of massive quiescent galaxies could be bridged by these objects, if the formation history inferred from the maximal stellar model is correct (Figure~\ref{fig:sfh}). A key question is whether the number densities of these objects match the number densities of massive quiescent galaxies at later times. Answering this question requires a larger area and/or a more well-defined selection function; for now we simply note that the number densities of these objects are comfortably below the number density of $z=3$ compact quiescent cores \citep{vanDokkum2015}. 
The presence of Balmer breaks at $z=7-8$, indicating very early and/or rapid formation, is probably the most straightforward prediction from finding ``maximally old'' $z=2-5$ quiescent galaxies. While the number density of these objects may be low, they are indicative of a class of objects with early and rapid formation. In addition, the strong spectroscopic selection effect, which we do not attempt to account for in this work, suggests that the value of $\sim10^{-5}$~Mpc$^{-3}$ (Figure~\ref{fig:rho}) is a lower limit of this class of objects. 
This is a key and perhaps convincing argument that in at least some of these objects, the stellar mass could be high (i.e., the AGN emission contribution to the continuum in the minimal/medium stellar models may be overestimated).

Another distinguishing property of the class of low-$z$ massive quiescent galaxies is the remarkable compactness and high stellar densities, with effective radii of 1~kpc or less at $z \sim 2$ \citep{vanDokkum08}. 
\citet{Baggen2023} showed that a fully stellar interpretation as presented in L23 (or similarly, the maximal stellar model of this paper) yields similar stellar densities to the cores observed at later times, albeit with significantly smaller sizes, by a factor of 10.

In addition, the enhanced SFR inferred from the maximal stellar model would corroborate to the surprising overabundance of luminous galaxies at $z \gtrsim 10$ (e.g., \citealt{Atek2023,Finkelstein2023,Robertson2023,Casey2023}). Their dense stellar structures would presumably be formed from very dense gas associated with highly efficient star formation \citep{Schmidt1959,Kennicutt1998}, perhaps in a similar manner as the super-star clusters with unusually high cloud surface densities in the local universe \citep{Smith2006,Turner2015}. 

Finally, the greater stellar mass case may be preferred on the grounds that it does not require a conveniently located AGN continuum to account for the observed spectral break. As the stellar mass decreases, so does the Balmer break strength in the galaxy spectrum, meaning that the AGN continuum must precisely match the shape and intensity needed to replicate the spectral break.

Therefore, we conclude that one possible interpretation is that the high-redshift Balmer break sample presented in this work are the progenitors of the first massive galaxies, observed directly after their rapid co-formation with their supermassive black holes. Certainly, this interpretation leaves difficult problems. 
We discuss the outstanding questions below, in connection with an alternative interpretation of these objects being low-mass galaxies hosting AGNs.

To start, it is not intuitive to explain the broad emission lines with a non-AGN origin, since stellar-driven scenarios are seemingly inconsistent with the lack of velocity offset between the narrow and broad emission line components, and the symmetry of the lines. The broad emission lines also suggest an abundance of ionizing photons, which is likewise difficult to make sense of under the stellar interpretation.
However, all the spectroscopically confirmed LRDs, which share similar but non-identical SED shapes as our sample, are under-luminous in X-ray \citep{Kocevski2023,Maiolino2023,Furtak2023:spec,Greene2023,Matthee2023,Wang2024:brd}. More recently, most LRDs are found to be under-detected in X-ray in Chandra observations \citep{Ananna2024,Yue2024}. It perhaps is then fair to speculate alternative, non-AGN-driven physical causes. 

Second, the implied formation is uncomfortably early and efficient, compared to the conventional assumption on the baryon to stellar conversion efficiency (Figure~\ref{fig:mhalo}; \citealt{Boylan-Kolchin2023}). Paradoxically, the cores of local massive galaxies appear to have bottom-heavy stellar initial mass functions (e.g., \citealt{conroy12}), with some evidence that this persists or even strengthens at $z \sim 2$ (\citealt{vanDokkum2024}, though see \citealt{Mercier2023} for an alternate take). This would increase the inferred stellar masses (without changing the observed SED in any way; \citealt{Wang2024:sys}) by a factor of a few, further increasing the tension with the cosmic baryon fraction. 

Third, the objects of this work are remarkably compact ($r_{\rm e} \lesssim 0.1$~kpc; \citealt{Baggen2023}). This is even more striking when comparing to \ruby QG-1, which has $r_{\rm e} \sim 0.6$~kpc \citep{deGraaff2024}, and the $z \sim 2-3$ compact quiescent galaxies, which are likewise much larger (by a factor of $\sim$10; \citealt{vanDokkum2015}). At face value, the stellar bodies of these objects would have to rapidly expand over the subsequent few hundred million years. Additionally, it is difficult to imagine a massive galaxy being less than 100~pc in size. This would indicate an increased importance for dynamical evolution effects normally reserved for dense globular clusters \citep{Spitzer1987,Vesperini1997}, e.g., segregation by mass, where the massive stars and binaries tend to sink toward the cores, and the low-mass stars move outward into the halo. Interestingly, the mass segregation, if it happened, would change the initial mass distribution, and thus relieve the tension in inferring uncomfortably large stellar mass in the current models, without evoking a change in the form of the initial mass function.

While thus far we have opted to center our discussion on the intriguing implications of the maximal stellar model---a choice motivated by the newly discovered high-$z$ Balmer breaks in this paper---we additionally acknowledge another possibility here.
An alternative interpretation, suggested by the minimal/medium stellar mass models, posits that these objects are low-mass galaxies hosting AGNs. Possible evolutionary tracks for this scenario have been discussed extensively in the literature (e.g., \citealt{Maiolino2023,Greene2023}). 
It is worthy emphasizing, however, that these objects need not necessarily all belong to the same category---the continuum composition may vary on an individual basis.
Future deeper, and redder observations are critical to distinguishing between galaxy and AGN contributions to the continuum in this population. 
We elaborate on the lingering questions affecting the interpretation of our sample, along with possible ways forward, in the section below.

\subsection{Key Remaining Questions About the Nature of the Balmer Break Sample}

One remaining puzzle is the contrast between the broad Balmer lines and the narrow forbidden emission lines. While the line widths have been shown to correlate well with mass in statistical samples \citep{Wuyts2016}, narrow line widths in \ha\ and CO have also been observed in some massive galaxies at $z\sim2.3$ when there is considerably less ambiguity about the stellar masses \citep{vanDokkum2015,mowla19}. Possibly, these narrow line widths can be explained by some combination of face-on disks and peculiar gas geometry (i.e., the emitting gas is not associated with the deep stellar potential well). The discrepancy between the significantly lower dynamical mass compared to the inferred stellar mass from the medium/maximal stellar SED models of all three objects, coupled with the absence of available evidence indicating the existence of disks within them, make such an argument less satisfying. However, it remains possible that the color-based selection function selects for a particular face-on orientation angle.

As for the $M_\star-M_{\rm BH}$ relation, we cautioned about the selection bias \citep{Lauer2007} in Section~\ref{subsec:bh}. Recently it has been argued that over-massive black holes are not inconsistent with the local relation after taking in account the selection bias \citep{Li2024}. 
A well-defined selection function was one of the key pieces in designing the observation strategy for the RUBIES program, and we will perform a more complete population-level analysis in a future paper.

A key missing piece of evidence is deep MIRI imaging---the stellar-dominated and AGN-dominated models can diverge at these wavelengths by a factor of $\gtrsim 2$. The MIRI filters probing the $1.6~\mu$m stellar bump \citep{Laurent2000,Sawicki2002}, and at the reddest end generally provide the most discriminating power. At longer wavelengths, stellar light is expected to decline at longer wavelengths, whereas AGN emission is expected to show a flat spectrum (if deficient in hot dust; \citealt{Williams2023,Perez-Gonzalez2024,Wang2024:brd}) or a rising spectrum (from a dusty torus). For instance, MIRI/F1280W happens to capture the peak of the $1.6~\mu$m stellar bump in the spectra of \ruby49140 and 55604, where the fluxes from the maximal and the minimal stellar models differ by $ \sim 3$. X-ray detections or firmer upper limits would also help to establish the AGN nature of these objects.

Finally, a word of caution: despite our exploration of several models, ranging from the minimal to maximal extremes of stellar mass, none of these models fully align with our current understanding of galaxy formation and evolution. This challenge arises from a combination of the difficulty in separating AGN and host galaxy light, and the ongoing debate surrounding evolutionary scenarios sparked by JWST's discoveries of (potentially) over-massive black holes and massive quiescent systems.
In light of all the complexities, the statements in this paper should all be considered contingent. There certainly remains ample room for reinterpretation in the future.

\subsection{Final Remarks}

JWST is revolutionizing our knowledge of the formation of early galaxies and their black holes. In this paper, we report a remarkable discovery of prominent Balmer breaks as early as $\zspec = 8.35$, and intriguingly, its coexistence with broad emission lines. 
The high-redshift Balmer breaks reveal unambiguously the presence of evolved stellar populations with extended formation histories within the first 600--800 Myr after the Big Bang. However, all of the examined explanations on the potential origins and evolutionary tracks for these objects leave key lingering questions.
Deeper spectroscopic data revealing stellar absorption features, and JWST/MIRI data sampling the red continuum would further elucidate the nature of these intriguing objects.
We conclude by emphasizing that observed Balmer breaks establishing the existence of evolved stellar populations with extended formation histories, presented herein mark an important development in understanding the origins and evolution of galaxies and their central supermassive black holes.

\section*{Acknowledgments}
We thank the anonymous referee for the helpful comments. 
B.W. and J.L. acknowledge support from JWST-GO-04233.009-A.
The Cosmic Dawn Center is funded by the Danish National Research Foundation (DNRF) under grant \#140.
This research was supported by the International Space Science Institute (ISSI) in Bern, through ISSI International Team project \#562 (First Light at Cosmic Dawn: Exploiting the James Webb Space Telescope Revolution). This work is based in part on observations made with the NASA/ESA/CSA James Webb Space Telescope. The data were obtained from the Mikulski Archive for Space Telescopes at the Space Telescope Science Institute, which is operated by the Association of Universities for Research in Astronomy, Inc., under NASA contract NAS 5-03127 for JWST. 
Computations for this research were performed on the Pennsylvania State University's Institute for Computational and Data Sciences' Roar supercomputer. This publication made use of the NASA Astrophysical Data System for bibliographic information.

\facilities{HST (ACS, WFC3), JWST (NIRCam, NIRSpec).}
\software{
Astropy \citep{2013A&A...558A..33A,2018AJ....156..123A,2022ApJ...935..167A}, 
dynesty \citep{Speagle2020}, 
EAzY \citep{Brammer2008}, 
emcee \citep{emcee},
Matplotlib \citep{2007CSE.....9...90H}, 
msaexp \citep{Brammer2022}, 
msafit \citep{deGraaff2023},
NumPy \citep{2020Natur.585..357H}, 
Prospector \citep{Johnson2021},
Python-FSPS \citep{pyfsps2023}.
}

\section*{Appendix}
\counterwithin{figure}{section}
\counterwithin{table}{section}
\renewcommand{\thesection}{\Alph{section}}
\setcounter{section}{0}

\section{An Age Indicator for High-redshift Galaxies\label{app:d4000}}

The commonly used age indicator, $D_{4000}$ \citep{Bruzual1983,Balogh1999},  measures the 4000\,\AA\ break, which results from the blanket absorption of high energy radiation from metals in stellar atmospheres and a deficiency of hot, blue stars. 
In this paper, we introduce a new definition, spectral break strength, to quantify the Balmer break strength using two windows at [3620,3720]\,\AA\ and [4000,4100]\,\AA.
While \citet{Binggeli2019} proposed using the fluxes in [3400,3600]\,\AA\ and [4150,4250]\,\AA, the larger separation means that this definition is more sensitive to the overall slope of the spectrum.
The wavelength windows used in this paper, however, avoid contamination from strong nebular line emission and are close enough to minimize the impact of dust attenuation and the overall spectrum slope.

\begin{figure*} 
\gridline{
  \fig{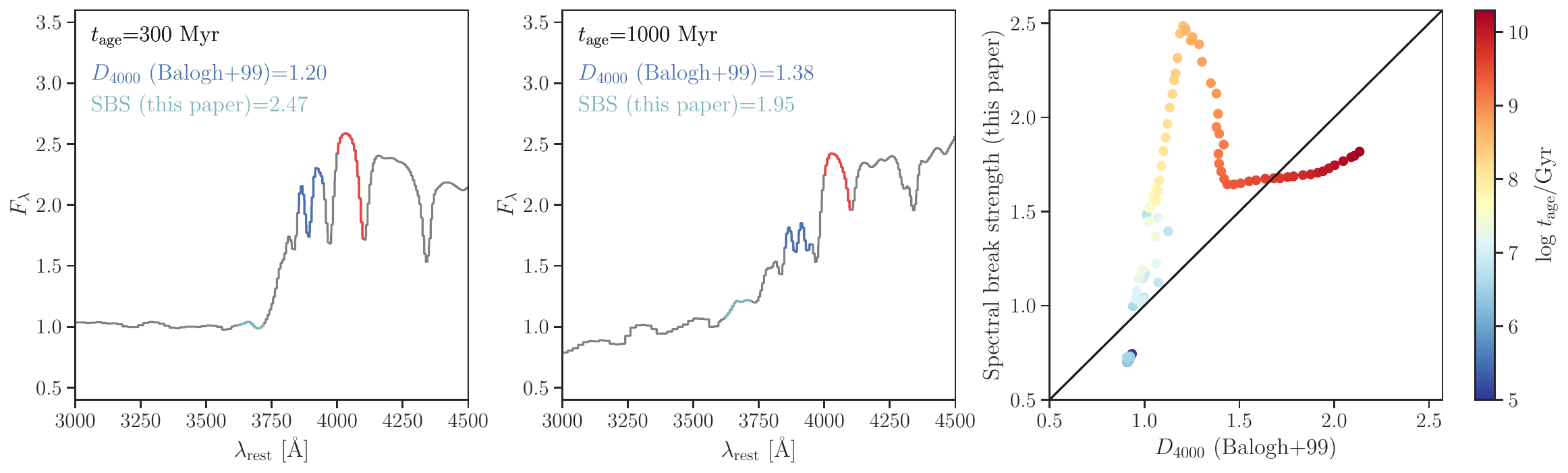}{0.95\textwidth}{}
}
\caption{Comparison between $D_{4000}$ \citep{Balogh1999} probing the 4000\,\AA\ break and the new spectral break strength (SBS) definition proposed in this paper, which is designed for the Balmer break. The two panels to the left show normalized model spectra at $t_{\rm age}=300$~Myr and 1000~Myr, respectively. The red color illustrates the wavelength window ([4000,4100]\,\AA) used by both \citet{Balogh1999} and this paper. Our definition differs in the bluer window. The blue color indicates the wavelength window ([3850,3950]\,\AA) used by \citet{Balogh1999}, whereas the cyan color indicates the wavelength window ([3620,3720]\,\AA) used by this paper. The values of the two age indicators are annotated in the upper left corner. The right panel contrasts the two definitions, color-coded by the age of the model spectra. The age-dependent variation is driven by the 4000\,\AA\ break being less visible than the Balmer break in galaxies with ages $\lesssim 1.5$~Gyr.}
\label{fig:app:d4000}
\end{figure*}

To gain more intuition on our new definition, we measure the $D_{4000}$ as defined in \citet{Balogh1999} on a set of model spectra, drawn from \texttt{FSPS} \citep{Conroy2010}. We assume solar metallicity, and exclude nebular emission lines. We also smooth the spectra to the JWST/Prism resolution.
The comparison, illustrated in Figure~\ref{fig:app:d4000}, shows that the difference is age-dependent, and differ mostly for galaxies with ages $\lesssim 1.5$~Gyr. This is expected, as the Balmer break is more prominent in these younger galaxies than the 4000\,\AA\ break. Our new age indicator is thus particularly suited for the high-redshift universe. 

\section{Inferred Spectral Break Strengths\label{app:prior}}

\begin{figure*}
\gridline{
  \fig{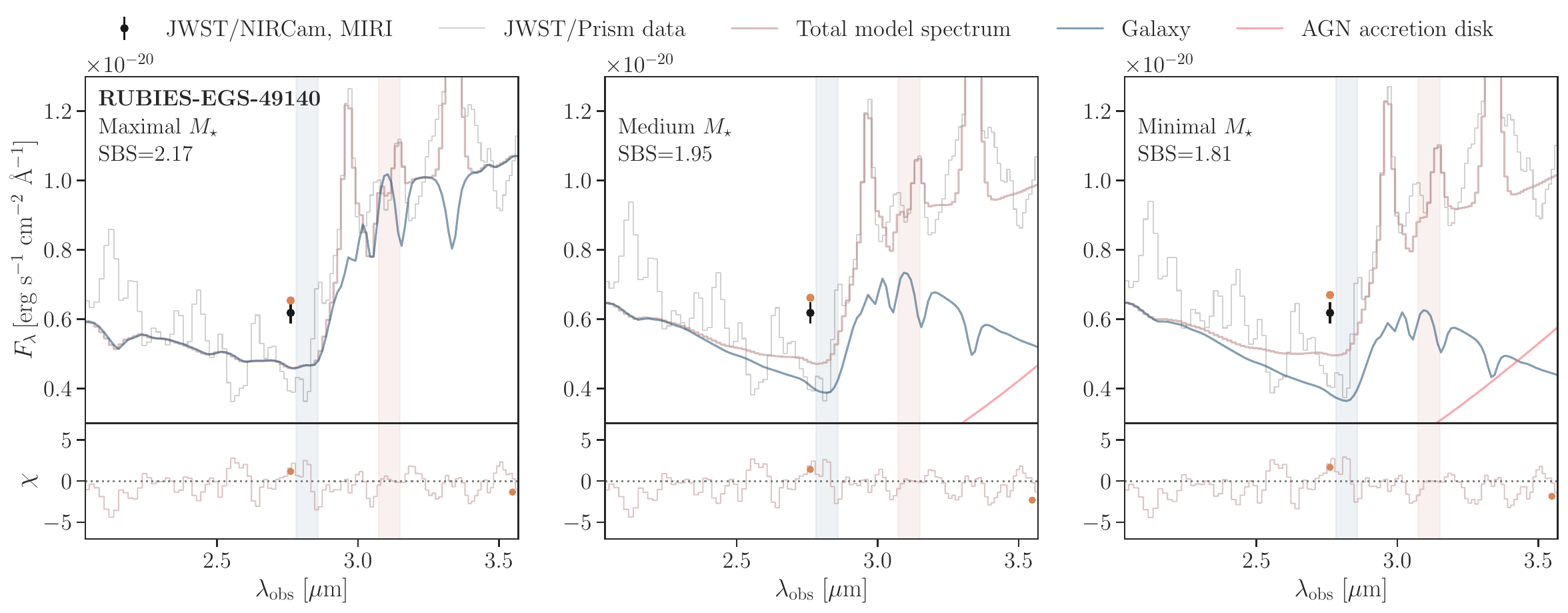}{0.9\textwidth}{}
}
\gridline{
  \fig{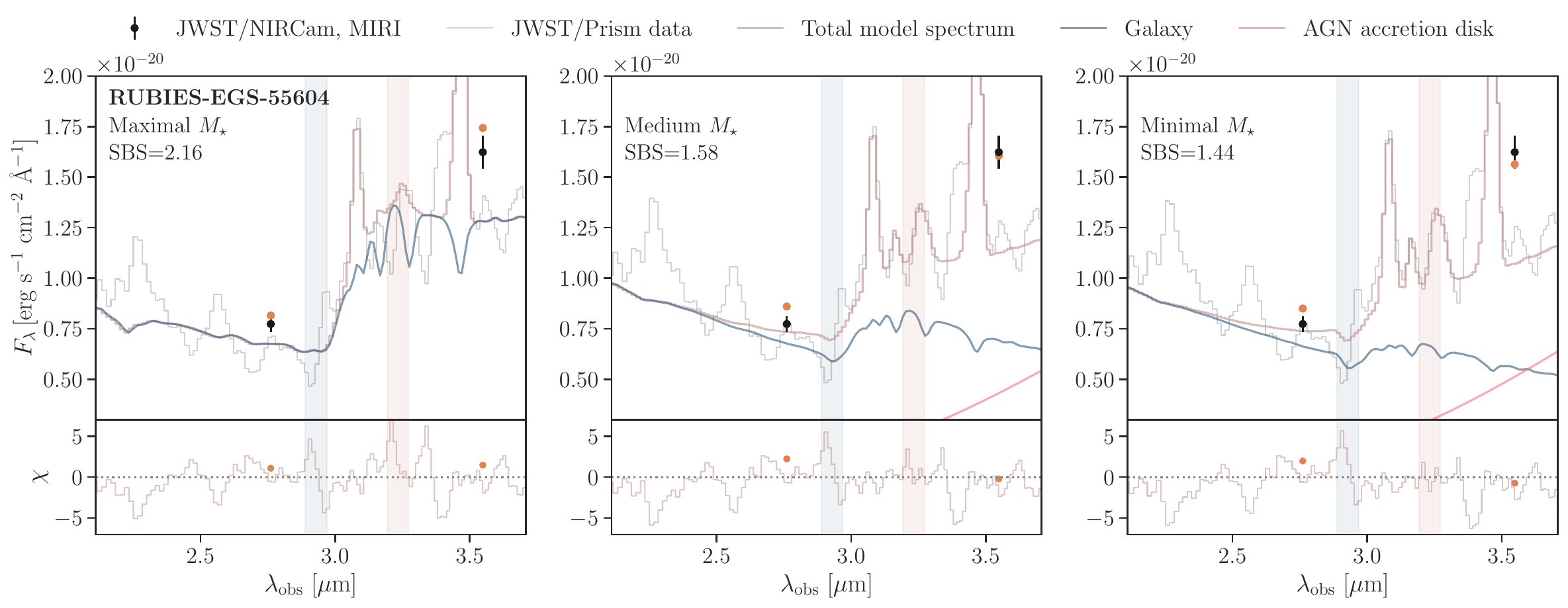}{0.9\textwidth}{}
}
\gridline{
  \fig{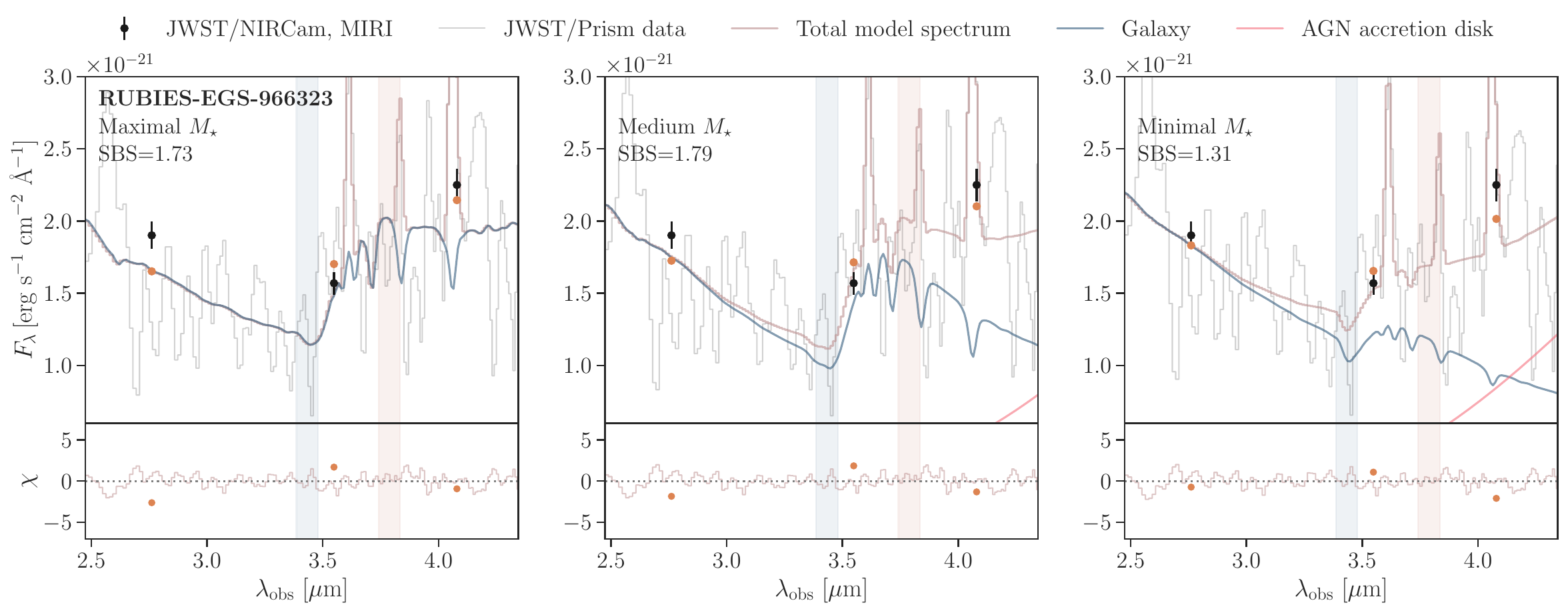}{0.9\textwidth}{}
}
\caption{Zoom-in of the spectral break region. In each row, we show the regions near the spectral break assuming the maximal, medium, and minimal stellar mass model from left to right. The residuals show minor differences, suggesting that all models provide a statistically acceptable solution. The spectral break strength (SBS), calculated as the flux ratio in the two wavelength windows in the blue and red shading, is indicated in the upper left corner in each panel. The minimal stellar mass model always predicts a weaker break strength than the data suggest, implying that the inferred age of the stellar population can be approximately taken as a lower limit.}
\label{fig:app:prior}
\end{figure*}

As a test for the constraining power of the data on the formation history, we examine the residuals around the spectral break resulting from the maximal, medium, and minimal stellar mass models, and compare the corresponding spectral break strength measured on the total model spectra (as opposed to on the galaxy spectra as is done in Section~\ref{sec:prosp}). As seen from Figure~\ref{fig:app:prior}, the minimal stellar mass model predicts a weaker break strength than the data suggest in all cases, implying that the data are sufficiently constraining for the purpose of inferring an extended formation history.

That being said, estimating a tight lower limit on the age of the stellar population is difficult at the current stage. First, the available data, as demonstrated throughout this paper, cannot break the degeneracies among the stellar populations, black hole properties, and dust attenuation. Second, while we have developed models that can describe the observed UV--optical SED, a full understanding of the underlying physical picture is still lacking. The most promising prospect may come from building a complete multi-wavelength view, leveraging the instrumental capability of e.g., Chandra, JWST/MIRI, and ALMA.

\bibliography{rubies_balmer_breaks_wang.bib}

\end{document}